\documentstyle[epsf,aps,prb,multicol]{revtex}
\begin{document}
%\psdraft
\draft
\title{\bf Quantum dot self consistent electronic structure
and the Coulomb blockade}
\author{M. Stopa} 
\address{RIKEN (The Institute of Physical
and Chemical Research) \\ 
2-1, Hirosawa, Wako-Shi
\\ Saitama 351-01, Japan \\ e-mail stopa@sisyphus.riken.go.jp}
\date{\today}
\maketitle
\begin{multicols}{2}
[\begin{abstract}
We  employ density functional theory to calculate the self
consistent  electronic  structure,  free energy and linear  source-drain
conductance of a lateral semiconductor quantum dot patterned via surface
gates  on  the  2DEG  formed  at  the   interface  of  a   $GaAs-AlGaAs$
heterostructure.  The  Schr\"{o}dinger   
equation  is reduced from  3D  to
multi-component  2D and solved via an eigenfunction
expansion in the dot.  This  permits the  solution of
the electronic structure for dot electron number $N \sim 100$.  
We present details of our derivation of the total
dot-lead-gates interacting free
energy in terms of the electronic structure results, which
is free of  capacitance  parameters. Statistical properties 
of the dot level spacings and 
connection coefficients to the leads are computed in the
presence of varying degrees of order in the donor layer. Based on
the  self-consistently  computed  free  energy  as a  function  of gate
voltages, $V_i$, and N, we modify the semi-classical  expression for the
tunneling conductance as a function of gate voltage  
through the dot in the linear
source-drain,   Coulomb  blockade  regime.
Among the many results presented, we demonstrate the existence of a 
shell structure in the dot levels which (a) results in envelope
modulation of Coulomb oscillation peak heights, (b) which
influences the dot capacitances and should be observable
in terms of variations in the activation energy for
conductance in a Coulomb oscillation minimum,
and (c) which possibly
contributes to departure of recent experimental results
from the predictions of random matrix theory.
\end{abstract}
]

%\narrowtext

\section{Introduction}

Study of the Coulomb  blockade  and  charging  effects in the  transport
properties   of   semiconductor   systems  is  peculiarly   suitable  to
investigation  through self-consistent electronic structure  techniques.
While the orthodox theory \cite{Lik}, in parameterizing the energy of
the system in terms of capacitances, is strongly applicable to metal
systems, the much larger ratio of Fermi wavelength to system size,
$\lambda_F / L$, in mesoscopic semiconductor devices, requires
investigation of the interplay of quantum mechanics and charging. 

In the first step beyond the orthodox theory, the ``constant 
interaction'' model of the Coulomb blockade supplemented the 
capacitance parameters, which were retained to characterize the gross
electrostatic contributions to the energy, with non-interacting quantum
levels of the dots and leads of the mesoscopic device \cite{Ruskies,Been}.
This theory was successful in explaining some of the fundamental features,
specifically the periodicity, of Coulomb oscillations in the conductance
of a source-dot-drain-gate system with varying gate voltage. Other
effects, however, such as variations in oscillation amplitudes, were not
explained.

In this paper we employ density functional (DF) theory to 
compute the self-consistently changing 
effective single particle levels of a lateral $GaAs-AlGaAs$
quantum dot, as a function of gate voltages,
temperature $T$, and dot electron number $N$ \cite{RComm}.
We also compute the total system free energy from the results of
the self-consistent calculation. We are then able to calculate 
the device conductance in the linear bias regime
without any adjustable parameters. Here we consider
only weak ($\stackrel{\sim}{<} 0.1 \; T$) magnetic fields
in order to study the effects of breaking time-reversal
symmetry.  We will present results for
the  edge state regime in  a  subsequent  publication
\cite{lp2}.

We include donor layer disorder in the calculation and present
results for the statistics of level spacings and partial
level widths due to tunneling to the leads. Recently we have employed
Monte-Carlo variable range hopping simulations to consider
the effect of Coulomb regulated ordering
of ions in the donor layer on the mode characteristics of split-gate
quantum {\it wires} \cite{BR2}. The results of those simulations
are here applied to quantum dot electronic structure.

A major innovation in this calculation is our method for
determining the two dimensional electron gas (2DEG) charge
density.
At each iteration of the self-consistent calculation, at
each point in the $x-y$ plane we determine the subbands $\epsilon_n (x,y)$ and 
wave functions $\xi^{xy}_n (z)$ in the $z$ (growth) direction. 
The full three dimensional density is then
determined by a solution of the multi-component 2D Schr\"{o}dinger
equation and/or 2D Thomas-Fermi approximation.

Among the many approximation in the calculation are the following.
We use the local density approximation (LDA) for exchange-correlation (XC),
specifically the parameterized form of Stern and Das Sarma \cite{SternDas}.
While the LDA is difficult to justify in small 
($N \sim 50-100$) quantum dots it is empirically known to give
good results in atomic and molecular systems where the density
is also changing appreciably on the scale of the Fermi wavelength
\cite{Slater}. 

In reducing the 3D Schr\"{o}dinger  equation to a multi-component  2D
equation we cutoff
the expansion in subbands, often taking only the 
lowest  subband  into  account. 
We also cutoff the wavefunctions by placing another artificial $AlGaAs$
interface at a certain depth (typically $200 \; \stackrel{0}{A}$)  away from
the first interface,  thereby ensuring the existence of subbands at all points
in the $x-y$ plane.  Generally the subband
energy of this bare square well is much  smaller  than the  triangular
binding to the  interface in all but those regions which are very nearly
depleted.

The dot electron states in the zero magnetic field regime
are simply  treated as spin  degenerate.  For $B \ne 0$ an
unrenormalized Land\'{e} g-factor of $-0.44$ is used.
We employ the  effective  mass
approximation  uncritically  and ignore the  effective  mass  difference
between  $GaAs$ and $AlGaAs$  ($m^* = 0.067 \; m_0$).  Similarly we take
the background  dielectric constant to be that of pure $GaAs$ ($\kappa =
12.5$) thereby ignoring image effects (in the $AlGaAs$).  
We ignore interface  grading and
treat the interface as a sharp  potential step.  These effects have been
treated in other  calculations of self-consistent  electronic  structure
for $GaAs-AlGaAs$ devices  \cite{SternDas} and have generally been found
to be small.

We mostly employ effective atomic units wherein 
$1 \; Ry^* = m^* e^4/2 \hbar^2 \kappa^2  \approx 5.8 \; meV$ and
$1 \; a_B^* = \hbar^2 \kappa/m^* e^2 \approx 100 \; \stackrel{0}{A}$.

The structure of the paper is as follows. 
In section II we first discuss the calculation of the electronic structure,
focusing on those features which are new to our method. Further subsections
then consider the treatment of discrete
ion charge and disorder, calculation of the total
dot free energy from the self-consistent electronic structure results,
calculation of the source-dot-drain conductance in the linear
regime and calculation of the dot capacitance matrix.
Section III provides new results which are further subdivided 
into basic electrostatic properties, properties of the
effective single electron spectra, statistics of level
spacings and widths and conductance in the Coulomb oscillation
regime. Section IV summarizes the principal conclusions
which we derive from the calculations.

\section{Calculations}

\subsection{Quantum dot self-consistent electronic structure}

\begin{figure}[hbt]
\setlength{\parindent}{0.0in}
 \begin{minipage}{\linewidth}
\epsfxsize=8cm
\epsfbox{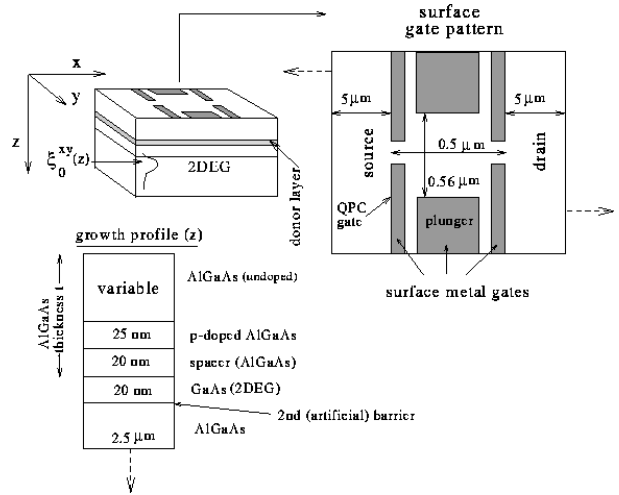}
\vspace*{3mm}
\caption{Schematic of device used in calculation. The 
$z$-subband structure throughout the plane are
calculated at each iteration of the self-consistency loop. Most
results presented with gate variation assume that both
the upper and lower pins of the relevant gate are simultaneously
varied. \label{fig1} }
 \end{minipage}
\end{figure}

We consider a 
lateral quantum dot patterned on a 2DEG heterojunction via metallic surface 
gates (Fig. \ref{fig1}). At a semiclassical level, other gate
geometries, such as a simple point contact or a multiple dot system, can be
treated with the same method \cite{BR2,G-res}. However, a 
full 3D solution of Schr\"{o}dinger's equation, even employing our subband 

expansion procedure for the $z$ direction, 
is only tractable in the current method when a region with a small
number of electrons ($N \le 100$) is quantum mechanically isolated, such as
in a quantum dot.

\subsubsection{Poisson equation and Newton's method}

In principal, a self-consistent solution is obtained by iterating the
solution of Poisson's equation and {\it some} method for calculating the 
charge density (see following sections II.A.2 and II.A.3). In practise, we follow 
Kumar {\it et al} \cite{Kumar}
and use an ${\cal N}$-dimensional Newton's method for finding the 
zeroes of the functional 
$\vec{F}(\vec{\phi}) \equiv {\bf \Delta} \cdot \vec{\phi} + \vec{\rho}(\vec{\phi})
+ \vec{q}$; where the potential, $\phi_i$, and density,
$\rho_i$, on the ${\cal N}$ discrete lattice
sites (${\cal N} \sim 100,000$) are written as vectors, $\vec{\phi}$ 
and $\vec{\rho}$.
The vector $\vec{q}$ represents the inhomogeneous contribution from
any Dirichlet boundary conditions, ${\bf \Delta}$ is the Laplacian
(note that here it is a matrix, not a differential operator), modified
for boundary conditions. Innovations for treating the Jacobian 
$\partial \rho_i / \partial \phi_j$
beyond 3D Thomas-Fermi,
and for rapidly evaluating the mixing parameter~$t$ (see Ref. \cite{Kumar})
are discussed below.

The Poisson grid spans a rectangular solid and hence the boundary conditions
on six surfaces must be supplied. Wide regions of the source and drain
must be included in order to apply Neumann boundary conditions
on these ($x = $ constant) interfaces, so a non-uniform mesh is essential. 
It is also possible to apply Dirichlet boundary conditions on these interfaces
using the ungated wafer (one dimensional) potential profile calculated 
off-line \cite{AFS}.
In this case, failure to include sufficiently wide lead regions shows
up as induced charge on these surfaces (non-vanishing electric field).
To keep the total induced charge on all surfaces below $0.5$ electron, 
lead regions of $\sim 5 \; \mu m$
are necessary, assuming a surface gate to 2DEG
distance (i.e. $AlGaAs$ thickness) of $1000 \stackrel{0}{A}$. In other words 
we need an aspect ratio of $50:1$.
We note that we ignore background compensation and merely
assume that the Fermi level is pinned at some fixed depth 
(``$z_{\infty}$'' $\sim 2.5 \; \mu m$)
into the $GaAs$ at the donor level.
The donor energy for $GaAs$ is taken as $1 \; Ry^*$ below
the conduction band. In the source and drain regions,
the potential of the 2DEG Fermi surface is fixed by the desired 
(input) lead voltage.

We apply Neumann boundary conditions at the $y = $ constant surfaces. 
The $z=0$ surface of
the device has Dirichlet conditions on the gated regions (voltage equal to the
relevant desired gate voltage) and Neumann conditions, 
$\partial \phi / \partial n = 0$,
elsewhere. 
This is equivalent to the ``frozen surface'' approximation
of \cite{JHD2}, further assuming a high dielectric constant for
the semiconductor relative to air. Further discussion of this
semiconductor-air boundary condition can be found in Ref. \cite{JHD2}.

\subsubsection{Charge density, quasi-2D treatment}

The charge density {\it within} the Poisson grid (i.e. not surface
gate charge) includes the 2DEG electrons and the ions in the
donor layer.
The treatment of discreteness, order and disorder in the donor ionic 
charge $\vec{\rho}_{ion}$ has been discussed in Ref. \cite{BR2}
in regards to quantum wire electronic structure. 
Some further relevant remarks are made below in section II.B.

As noted above, we take advantage of the quasi-2D nature of the 
electrons at the $GaAs-AlGaAs$ interface to simplify the calculation for 
their contribution to
the total charge. Given $\vec{\phi}$, we begin by solving Schr\"{o}dinger's 
Eq. in the $z$-direction {\it at every point} in the $x-y$ plane,
\begin{equation}
[-\frac{\partial^2}{\partial z^2} + V_B (z) + e \phi(x,y,z)] \xi^{xy}_n (z)
= \epsilon_n (x,y) \xi^{xy}_n (z) \label{eq:eqz}
\end{equation}
where $V_B(z)$ is the potential due to the conduction band offset between
$GaAs$ and $Al_x Ga_{1-x} As$. We generally employ fast Fourier
transform with $16$ or $32$ subbands. 

In order that there be a discrete spectrum at each point in the
$x-y$ plane, it is convenient to take $V_B(z)$ as a {\it square well} potential
(Fig. \ref{fig1}). That is, we effectively cutoff the wave 
function with a second barrier,
typically $200 \stackrel{0}{A}$ from the primary interface. In undepleted
regions the potential is still basically triangular and only the tail of 
the wave function is affected. However, near the border between depleted and 
undepleted regions the artificial second barrier will introduce some error
into the electron density. This is because 
as a depletion region is approached, the binding {\it electric
field} at the 2DEG interface (slope of the triangular potential)
reduces, in addition to the interface potential itself rising.
Consequently, all subbands become degenerate and {\it near the edge
electrons are three dimensional} \cite{McEuenrecent}. We have checked that this
departure from interface confinement, and in general in-plane
gradients of $\xi^{x,y}_n (z)$
contribute negligibly to quantum dot level energies. 
However, theoretical descriptions of 2DEG edges commonly assume perfect 
confinement of electrons in a plane.
In particular the description of edge excitations in the quantum Hall
effect regime in terms of a chiral Luttinger liquid
\cite{Wen} may be complicated in real samples by the emergence
of this vanishing energy scale and collective modes related to it.

Assuming only a single $z$-subband now and dropping the index $n$,
we determine the
charge distribution in the $x-y$ plane from the effective potential 
$\epsilon (x,y)$,
employing a 2D Thomas-Fermi approximation for the charge in the leads and
solving a 2D Schr\"{o}dinger equation in the dot. In order that the dot 
states
be well defined, the QPC saddle points must be classically inaccessible. 
(If this
is not the case it is still possible to use a Thomas-Fermi approximation 
throughout the
plane for the charge density \cite{BR2,G-res}).
In the dot, the density is determined from the eigenstates by
filling states according to a Fermi distribution
either to a prescribed ``quasi-Fermi energy'' of the dot, or to a fixed number
of electrons. It has been pointed out that a Fermi distribution for the level 
occupancies in the dot is an inaccurate approximation to the correct grand
canonical ensemble distribution \cite{Been}. Nonetheless, for small dots
($N \stackrel{<}{\sim} 15$)
Jovanovic {\it et al.} \cite{Jovanovic} have shown that, regarding the
filling factor,  the discrepancy
between a Fermi function evaluation and that of the full grand canonical 
ensemble 
is $\sim 5\%$ at half filling and significantly smaller away from 
the Fermi surface. As $N$ increases the discrepancy should become smaller.

\subsubsection{Solution of Schr\"{o}dinger's equation in the dot}

To solve the effective 2D Schr\"{o}dinger's equation in the dot,
\begin{equation}
(-{\bf \nabla}^2 + \epsilon({\bf x}) ) f ({\bf x}) = E f ({\bf x}) \label{eq:eqE}
\end{equation}
we set the 2D potentials throughout the {\it leads} to their values at the 
saddle points, 
thereby ensuring that
the wave functions decay uniformly into the leads. Thus the energy of the higher
lying states will be shifted upward slightly. 
In seeking a basis in which to expand the solution of Eq. \ref{eq:eqE} we
must consider the approximate shape of the potential. The quantum dots which we 
model here
are lithographically approximately square in shape. However 
the potential at the 2DEG
level and also the effective 2-D potential $\epsilon(r,\theta)$, (now in 
polar coordinates) are to lowest
order azimuthally symmetric. The {\it radial} dependence of the potential is 
weakly parabolic across the center. 
Near the perimeter higher order terms become important
(cf. figure \ref{fig3}b and Eq. \ref{eq:phi}).

As the choice of a good basis is not completely clear, we have tried two 
different sets of functions: Bessel functions and the so-called Darwin-Fock (DF) 
states \cite{Darwin}.
The details of the solution for the eigenfunctions and eigenvalues differ
significantly whether we use the Bessel functions or the DF states.
The Bessel function case is largely numerical whereas the DF functions together
with polynomial fitting of the azimuthally symmetric part of
the radial potential allow a considerable amount
of the work to be done analytically. Further, neither of the two bases comes
particularly close to fitting the somewhat eccentric shape of the actual dot
potential. It is therefore gratifying that comparing the eigenvalues determined
from the two bases when reasonable cutoffs are used, we find for up to the
$50^{th}$ eigenenergy agreement to three significant 
figures, or to within roughly $5 \; micro \; eV$.

\subsubsection{Summary and efficiency}

To summarize the calculation, we begin by choosing the device dimensions
such as the gate  pattern,  the  ionized  donor  charge  density and its
location relative to the 2DEG, the aluminum concentration for the height
of the barrier, and the  thickness of the $AlGaAs$  layer.  We construct
non-uniform grids in $x$, $y$ and $z$ that best fit the 
device within a total of about $10^5$ points. 
Gate  voltages, temperature, source-drain  voltages, and
either the  electron  number $N$ or the  quasi-Fermi  energy of the dot
are inputs.  The iteration  scheme begins 
with a guess of  $\vec{\phi}^{(0)}$.  The
1-D  Schr\"{o}dinger  equation  is solved at each  point in the  $x-y$  plane  and
an effective 2-D potential  $\epsilon(x,y)$ for one or
at most two subbands is thereby determined.  
Taking $|\xi^{xy}_n (z)|^2$ for the $z$-dependence of the charge density, 
we compute the 2D dependence in the leads using
a 2D Thomas-Fermi approximation and in the dot by solving Schr\"{o}dinger's  equation 
and filling the computed  states according to a Fermi distribution. We 
compute $\vec{F} (\vec{\phi}^{(0)})$,
which is a measure of how far we are from self-consistency, and solve for 
$\delta \vec{\phi}$, the potential increment, using a mixing parameter $t$.
This gives the next estimate for the potential $\vec{\phi}^{(1)}$. The 
procedure is iterated and convergence is
gauged by the norm of $F$.

In practise there are many tricks which one uses to hasten (or
even obtain !) convergence. First, we use a scheme 
developed by Bank and Rose \cite{Bank,Kumar} to search for an
optimal mixing parameter $t$.
Repeated calculation of Schr\"{o}dinger's equation, which is very
costly, is in principle required in the search for $t$. 
Far from convergence the Thomas-Fermi approximation can be used
in the dot as well as the leads. Nearer to convergence we find that
diagonalizing $t \; \delta \vec{\phi}$ in a basis of
about ten states near the Fermi surface, treating the charge in the other
filled states as inert, is highly efficient.
Periodically the full solution of Schr\"{o}dinger's equation
is employed to update the wave functions. 

The wave function information is also used to make a better estimate
of $\partial \rho_i / \partial \phi_i$. The 3D Thomas-Fermi method
for estimating this quantity does not account for the fact that
the change in density at a given grid point will be most strongly
influenced by the changes in the occupancies of the partially
filled states at the Fermi surface. Thus use of these wave functions
greatly improves the speed of the calculation.

\subsection{Disorder}

Evidence of Coulombic {\it ordering}
of the donor charge in a modulation doping layer adjacent to a 2DEG 
has recently accumulated \cite{Buks}.
When the fraction ${\cal F}$ of ionized donors among all donors is less
than unity, redistribution of the ionized sites through hopping can lead
to ordering of the donor layer charge \cite{Efros,BR2}.  

In this paper we consider the effects of donor charge distribution on
the statistical properties of quantum dot level spectra, in particular the
unfolded level spacings, and on the connection coefficients to the 
leads $\Gamma_p$ of the individual states (see below). These dot
properties are calculated with ensembles of donor charge which
range from completely random (identical to ${\cal F}=1$, no ion
re-ordering possible) to highly ordered (${\cal F} \sim 1/10$).
For a discussion of the glass-like properties of the donor layer 
and the Monte-Carlo variable range hopping calculation which 
is used to generate ordered ion ensembles, see Refs. \cite{BR2} and 
\cite{ISQM2}.

Note that hopping is assumed to take place at temperatures ($\sim 160 \, K$) 
much higher than the sub-liquid Helium temperatures at which the dot electronic
structure is calculated. Thus the ionic charge distributions
generated in the Monte-Carlo calculation are, for
the purposes of the 2DEG electronic structure calculation,
considered fixed space charges which are specifically
not treated as being in thermal equilibrium
with the 2DEG.

The region where the donor charge can be taken as discrete is limited by
grid spacing and hence computation time. 
In the wide lead regions and wide region lateral to the
dot the donor charge
is always treated as ``jellium.'' Also,
to serve as a baseline, we calculate the
dot structure with jellium across the dot region as well.
We introduce the term ``quiet dot''
to denote this case.

\subsection{Free energy}

To calculate the total interacting free energy we 
begin from the 
semi-classical expression
\begin{eqnarray}
F(&\{n_p\},Q_i,V_i) = \sum_p n_p \varepsilon_p^0 + 
\frac{1}{2} \sum_i^M Q_i V_i \nonumber \\
& - \sum_{i \ne dot} \int dt \; V_i (t) I_i (t) \label{eq:cl}
\end{eqnarray}
where $n_p$ are the occupancies of non-interacting dot energy 
levels $\varepsilon_p^0$;
$Q_i$ and $V_i$ are the charges and voltages of the $M$ distinct 
``elements'' into which we divide the system: dot, leads and gates.
$I_i$ are the currents supplied by power supplies to the elements. 

The {\it self-consistent} energy levels for the electrons
in the dot are $\varepsilon_p = < \psi_p \mid - \nabla^2 + V_B (z) +e \phi ({\bf r})
\mid \psi_p >$. A sum over these levels double counts
the electron-electron interaction. Thus, for the terms in Eq. \ref{eq:cl} 
relating to the dot, we make the replacement:
\begin{eqnarray}
& \sum_p  n_p \varepsilon_p^0  + \frac{1}{2} Q_{dot} V_{dot} \rightarrow
\sum_p n_p \varepsilon_p \nonumber \\
& - \frac{1}{2} \int d{\bf r} \rho_{dot}({\bf r})
\phi ({\bf r}) + \frac{1}{2} \int d{\bf r} \rho_{ion}({\bf r})  \phi ({\bf r})
\end{eqnarray}
where $\rho_{dot}({\bf r})$ refers only to the charge in the dot
states and $\rho_{ion}({\bf r})$ refers to all the charge in the donor layer.

We have demonstrated \cite{BR1,MCO} that 
previous investigations \cite{Been,vanH} had failed to correctly include the
work from the power supplies, particularly to the source and drain leads,
in the energy balance for tunneling between leads and dot in the 
Coulomb blockade regime. Here, we assume a low impedance environment
which allows us to make the replacement:
\begin{equation}
\frac{1}{2} \sum_{i \ne dot} Q_i V_i - 
\sum_{i \ne dot} \int dt \; V_i (t) I_i (t) \rightarrow
- \frac{1}{2} \sum_{i \ne dot}  Q_i V_i.
\end{equation}
The charges on the gates are determined from the gradient of the potential 
at the various surface regions, the voltages being given. 
Including only the classical electrostatic energy of the leads,
the total free energy is \cite{RComm}:
\begin{eqnarray}
& F(\{n_p\},N,V_i)  = \sum_{p} n_p 
\varepsilon_{p} - \frac{1}{2} \int d{\bf r} 
\rho_{dot}({\bf r}) \phi ({\bf r}) \nonumber \\
& + \frac{1}{2} \int d{\bf r} \rho_{ion}({\bf r}) \phi ({\bf r})
- \frac{1}{2} \sum_{i \; \epsilon \; leads} \; \int d{\bf r} 
\rho_i ({\bf r}) \phi ({\bf r}) \nonumber \\
& - \frac{1}{2} \sum_{i \; \epsilon \; gates} Q_i V_i \label{eq:free}
\end{eqnarray}
where the energy levels, density, potential and induced charges are
implicitly functions of $N$ and the
applied gate voltages $V_i$. Note that the occupation number dependence of
these terms is ignored. In the $T=0$ limit the electrons occupy the 
lowest $N$ states of the dot, and the free energy is denoted
$F_0 (N,V_i)$.

\subsection{Conductance}

The master
equation formula for the linear source-drain conductance though the
dot, derived by several authors \cite{Been,Ruskies,Meir}
for the case of a fixed dot spectrum, is modified to the self-consistently
determined free energy case as follows \cite{RComm}:
\begin{eqnarray}
& G(V_g)  = \displaystyle{\frac{e^2}{k_B T} \sum_{\{ n_{i} \} }}
P_{eq}( \{ n_{i} \} )  \sum_{p} \delta_{n_{p},0} 
\displaystyle{\frac{\Gamma_p^s \Gamma_p^d}{\Gamma_p^s + 
\Gamma_p^d}} \nonumber \\
& \times f(F(\{n_i+p\},N+1,V_g)  
- F(\{n_i\},N,V_g) - \mu ) \label{eq:cond}
\end{eqnarray}
where the first sum is over dot level occupation configurations and the second
is over dot levels. The equilibrium probability distribution 
$P_{eq} ( \{ n_i \} )$ is
given by the Gibbs distribution,
\begin{equation}
P_{eq} ( \{ n_i \} ) = \frac{1}{Z} exp[- \beta (F(\{n_i\},N,V_g) - \mu)]
\end{equation}
and the partition function is
\begin{equation}
Z \equiv \sum_{\{ n_{i} \} } exp[- \beta (F( \{ n_i \} ,N,V_g) - \mu)]
\end{equation}
note that the sum on occupation configurations, $\{ n_{i} \}$, includes
implicitly a sum on $N$.
In Eq. \ref{eq:cond} $f$ is the Fermi function, $\mu$ is the
electrochemical potential of the source and drain and $\Gamma_p^{s(d)}$ 
are the elastic couplings of level $p$ to source (drain).
The notation $\{ n_i + p \}$ denotes
the set of occupancies $\{ n_i \}$ with the $p^{th}$ level, previously
empty by assumption, filled. In Eq. \ref{eq:cond} it is assumed that
only a single gate voltage, $V_g$ 
(the ``plunger gate'', cf. Fig. \ref{fig1}), is varied.

\subsection{Tunneling coefficients}

The elastic couplings in Eq. \ref{eq:cond} are calculated from the 
self-consistent wave functions \cite{Bardeen}:
\begin{equation}
\hbar \Gamma_{np} = 4 \kappa^2 W_n^2 (a,b) \; \left| \int dy \; 
f_p (x_b,y) \chi^*_n (x_b,y) \right|^2 \label{eq:tun}
\end{equation}
where $f_p (x_b,y)$ is the two dimensional part of the $p^{th}$ wave function 
evaluated at the midpoint of the barrier, $x_b$, and $\chi^*_n (x_b,y)$ 
is the $n^{th}$ channel wavefunction decaying into the barrier from the 
leads, $W_n(a,b)$ is the barrier penetration factor between
the classical turning point in the lead and the point $x_b$,
for channel $n$ computed in the WKB approximation, and $\kappa$ is the 
wave vector at the matching point. Though the channels are 1D we use 
the two dimensional density of states characteristic of the wide 
2DEG region \cite{Matveev}.

\subsection{Capacitance}

Quantum dot system electrostatic energies are commonly estimated on the
basis of a capacitance model \cite{various}. When the self-consistent level energies
and potential are known the total free energy can be computed without 
reference to capacitances. However, the widespread use of this model
and the ease with which capacitances can be calculated from our
self-consistent results (see below) encourages
a discussion. 

For a collection of $N$
metal elements with charges $Q_i$ and voltages $V_j$ 
the capacitance matrix, defined by \cite{BR1,meandYasu}
$Q_i = \sum_{j=1}^{N} C_{ij} V_j$, can be written in
terms of the Green's function $G_D ({\bf x,x^{\prime}})$
for Laplace's equation
satisfying Dirichlet boundary conditions
on the element surfaces:
\begin{equation}
C_{ij} = \frac{1}{4 \pi^2} \int d \Omega_i \int d \Omega_j
\hat{n}_j \cdot \vec{\nabla}_x (\hat{n}_i \cdot \vec{\nabla}_{x^\prime}
G_D ({\bf x,x^{\prime}}))
\end{equation}
where the integrals are over element surfaces with $\hat{n}_j$
the outward directed normal. 

In a system with an element of size $L$ not much greater than the screening 
length $\lambda_s$, the voltage of the component, and hence the 
capacitance, is not well defined \cite{meandYasu,Buttikercap}.
In this case, as discussed in reference \cite{meandYasu},
the capacitance can no longer be written in terms of the solution
of Poisson's equation alone, but must take account of the 
full self-consistent determination of the $i^{th}$ 
charge distribution $\rho_i({\bf x})$
from the $j^{th}$ potential $\phi_j({\bf x})$ $\forall i,j$. 
In general the capacitance will
then become a kernel in an integral relation. A relationship
of this kind has recently been derived in terms of the
Linhard screening function by B\"{u}ttiker \cite{Buttikercap}.

To compute the dot self-capacitance from the calculated self-consistent
electronic structure we have three separate procedures.
In all three cases we
vary the Fermi energy of the dot by some small amount to change
the net charge in the dot. This requires that the QPCs be closed.
For the first method
the total charge variation of the dot is divided by the 
change in the electrostatic potential minimum of
the dot. This is taken as the dot self-capacitance $C_{dd}$. 
A second procedure for the dot self-capacitance is to 
divide the change in the dot charge simply
by the fixed, imposed change of the Fermi energy. This result is 
denoted $C_{dd}^\prime$. Since the change in the potential
minimum of the dot is not always equal to the change of
the Fermi energy these results are not identical.
Finally, we can fit the computed free energy $F(N,V_g)$ to a
parabola in $N$ at each $V_g$. If the quadratic term is
$\alpha N^2$ then the final form for the self-capacitance is
$C_{dd}^{\prime \prime} = 1/(2 \alpha)$ (primes are {\it not}
derivatives here). This form, which
also serves as
a consistency check on our functional for the energy,
is generally quite close to the first form and
we present no results for it.

For the capacitances between dot and gates or leads, the extra
dot charge (produced by increasing the Fermi energy in the dot)
is screened in the gates and the leads so that the net
charge inside the system (including that on the gated boundaries)
remains zero. The fraction of the charge screened in a
particular element gives that element's capacitance
to the dot as a fraction of $C_{dd}$.

\section{Results}

We consider only a small subspace of the huge available 
parameter space. For the results presented here we have fixed
the nominal 2DEG density to $1.4 \times 10^{11} \; cm^{-2}$ and
the aluminum concentration of the barrier to $0.3$. The lithographic
gate pattern is shown in figure \ref{fig1}, as is the growth profile
(including our artificial second barrier). Some results are 
presented with a variation of the total thickness $t$ of the
AlGaAs (Fig. \ref{fig1}).

To interpret the results we note the following considerations.
Hohenberg-Kohn-Sham theory provides only that the ground
state energy of an interacting electron system can be
written as a functional of the density \cite{HKS1,HKS2}.
The single particle eigenvalues $\varepsilon_p$ have, strictly
speaking, no
physical meaning. However, as pointed out by Slater
\cite{Slater}, the usefulness of DF theory depends to 
some extent on being able to interpret the energies 
and wave functions as some kind of single particle 
spectrum. In the Coulomb blockade regime it is particularly
important to be clear what that interpretation, and its
limitations, are.

A distinction is commonly made between the addition
spectrum and the excitation spectrum for quantum dots
\cite{McEuen,Ashoori}. Differences between our effective
single particle eigenvalues represent an approximation
to the excitation spectrum. As a specific
example, in the
absence of depolarization and excitonic effects
the first single
particle excitation from the $N$-electron ground state
with gate voltages $V_i$
is $\varepsilon_{N+1}(N,V_i)-\varepsilon_{N}(N,V_i)$.

The addition spectrum, on the other hand,
depends on the energy difference between the
ground states of the dot {\it interacting
with its environment} at two different $N$.
Thus, in our formalism, the addition spectrum
is given by differences in $F(\{n_p\},N,V_i)$ at 
neighboring $N$, possibly further modulated by excitations,
i.e. differences
in the occupation numbers $\{ n_p \}$.

In contrast to experiment, the electronic structure can
be determined for arbitrary $N$ and $V_i$ (so long as
the dot is closed). This includes both non-integer
$N$ as well as values which are far from equilibrium
(differing chemical potential) with the leads. The
``resonance curve'' \cite{RComm} is given by the $N$
which minimizes $F_0 (N,V_g)$ at each $V_g$ (gates other than
the plunger gate are assumed fixed). This occurs when
the chemical potential of the
dot equals those of the leads (which are taken as equal
to one another and represent the energy zero)
and gives the most probable electron number. Results
presented below as a function of varying gate voltage, 
particularly the spectra in Figs. \ref{fig10} and \ref{fig14},
are assumed to be along the resonance curve. 

\subsection{Electrostatics}

\begin{figure}[hbt]
\setlength{\parindent}{0.0in}
 \begin{minipage}{\linewidth}
\epsfxsize=8cm
\epsfbox{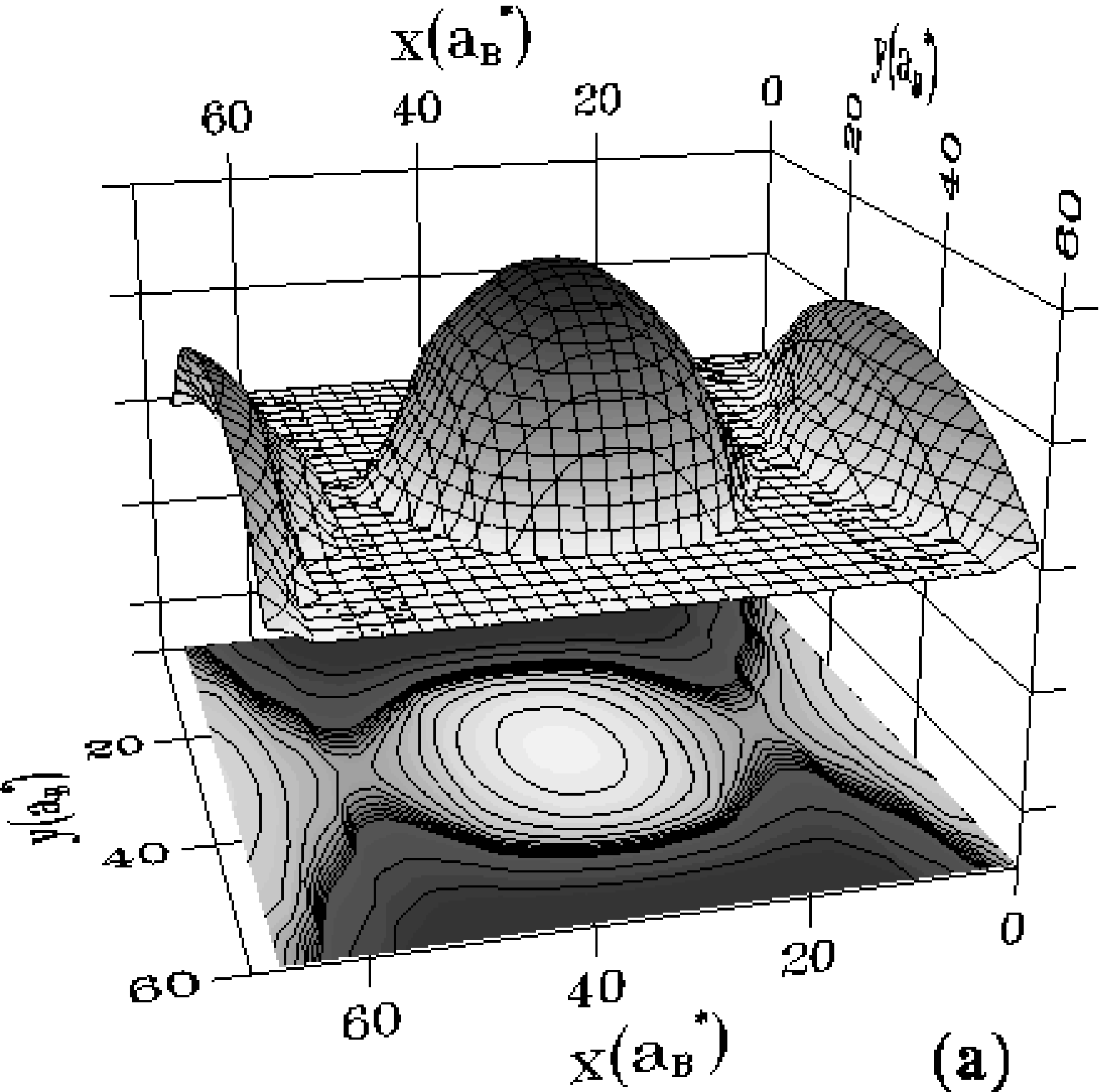}
\vspace*{3mm}
\epsfxsize=8cm
\epsfbox{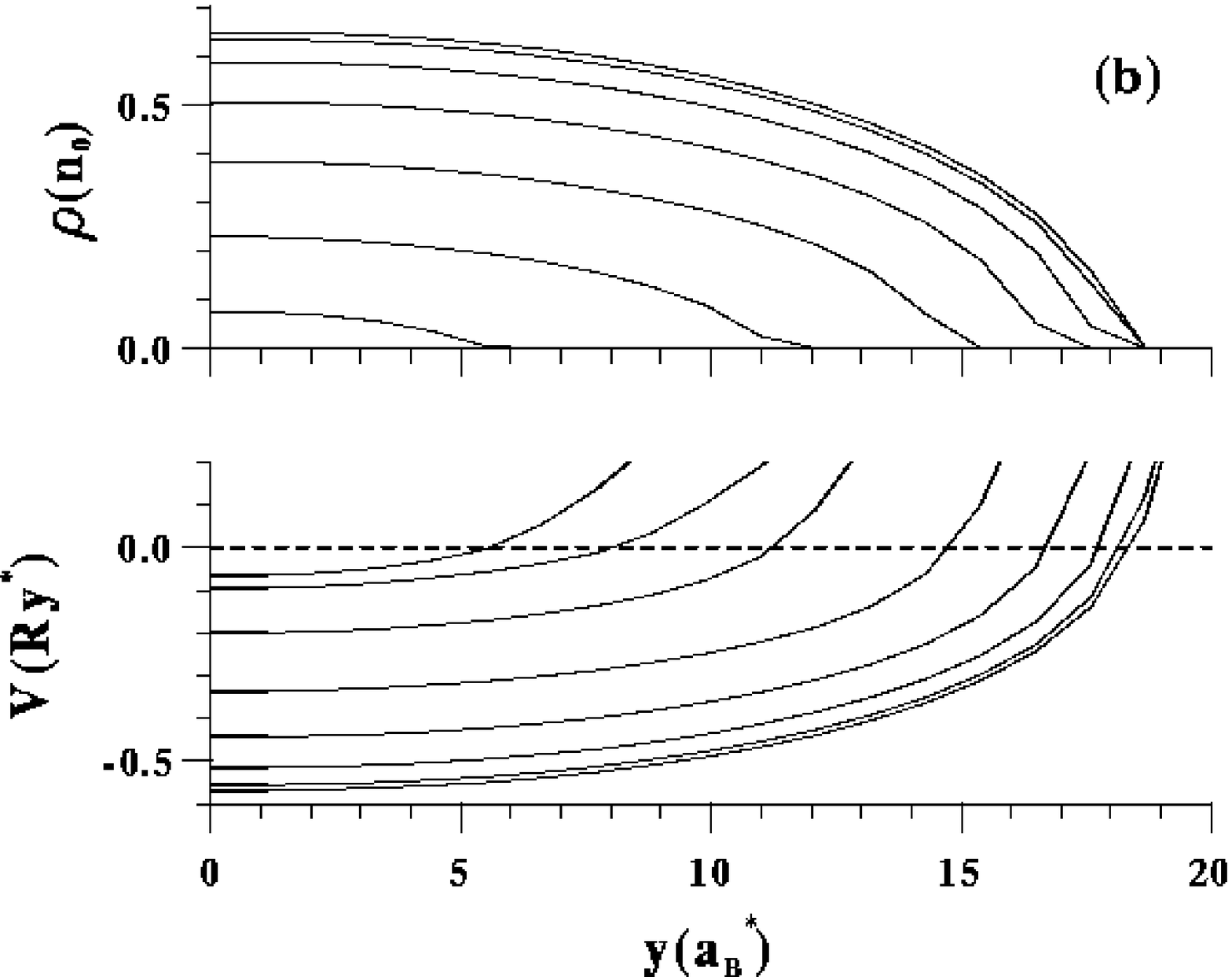}
\vspace*{3mm}
\caption{(a) Contour plot for density and potential, quiet dot, TF. 
Isolines in 
potential spaced at $\sim 0.1 \; Ry^*$ up to $0.5 \; Ry^*$
above Fermi level, after which much more widely. Density isoline
spacing $\sim 0.01 \; a_B^{* \, -2}$, maximum density 
$\sim 0.1 \; a_B^{* \, -2}$. Ripples near QPCs are finite grid size
effect; plotted $x-y$ mesh shows every other grid line. 
(b) Transverse (y-direction) half-profiles of density and potential 
corresponding to (a), taken at $3.3 \; a_B^*$ intervals
from dot center. Uppermost potential trace, entirely above Fermi surface,
is in QPC ($x \approx 54 \; a_B^*$ in Fig. 2a) where density is zero. 
Density is scaled to nominal
2DEG value $0.14 \; a_B^{* \, -2} \approx 1.4 \times 10^{11} \; cm^{-2}$.
\label{fig3} }
 \end{minipage}
\end{figure}

Figure \ref{fig3}a shows an example of a 
potential profile along with a corresponding
density plot for a quiet dot containing $62$ electrons. The basic 
potential/density configuration, as well as the
capacitances are highly robust. These data are computed completely in the 
2D Thomas-Fermi approximation, single $z$-subband, 
at $T = 0.1 \; K$. Solution of Schr\"{o}dinger's
equation or variation of $T$ result in only subtle changes. The depletion region 
spreading is roughly $100 \; nm$. Figure \ref{fig3}b shows a set of potential
and density profiles along the y-direction (transverse to the current direction)
in steps of $3.3 \; a_B^*$ in $x$, from the QPC saddle point to the dot center.
Note that the density at the dot center is only about $65 \%$ of the ungated 2DEG
density. Correspondingly the potential at the center is above the floor of the
ungated 2DEG ($\sim -0.9 \; Ry^*$).

We discuss a simple model for the potential shape of a
circular quantum dot below (Sec. III.B.1). Here we note only
that the radial potential can be regarded as parabolic to lowest
order with quartic and higher order corrections whose influence
increases near the perimeter. In Thomas-Fermi studies on larger 
dots \cite{MCO,G-res} with a comparable aspect ratio we find that
the potential and density achieve only $90 \; \%$ of their ungated 2DEG value 
nearly $200 \; nm$ from the gate. Regarding classical billiard calculations for
gated structures therefore \cite{chaos1,Been2,Bird,Ferry} even in the
absence of impurities it is difficult to see how the ``classical''
Hamiltonian at the 2DEG level can
be even approximately integrable unless the lithographic 
gate pattern is azimuthally symmetric \cite{square}. 

\begin{figure}[hbt]
\setlength{\parindent}{0.0in}
 \begin{minipage}{\linewidth}
\epsfxsize=8cm
\epsfbox{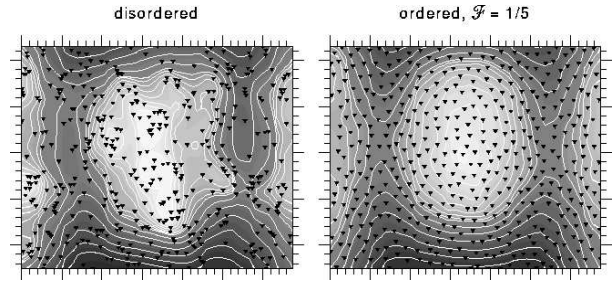}
\vspace*{3mm}
\caption{Contour plots of dot showing ion placement for disordered case
(left) and ordered (${\cal F}=1/5$) 
ion distribution, TF. Isolines at $0.08 \; Ry^*$ up to
Fermi surface, wider thereafter. Gate voltages and locations identical
in the two cases. Note particularly position of right QPC determined
by ions in disordered case. \label{fig4} }
 \end{minipage}
\end{figure}

The importance of the remote ionized impurity distribution is demonstrated in
figure \ref{fig4} which shows a quantum dot with randomly placed ionized 
donors on
the left and with ions which have been allowed to reach quasi-equilibrium via
variable range hopping, on the right. In both cases the total ion number in
the area of the dot is fixed. The example shown here for
the ordered case assumes, in the variable range hopping calculation,
one ion for every five donors (${\cal F}=1/5$). As in Ref. \cite{BR2}
we have, for simplicity, ignored the negative $U$ model for the donor
impurities (DX centers), which is still controversial 
\cite{Buks,Mooney,Yamaguchi}.
If the negative $U$ model, at some barrier aluminum concentration,
is correct, the most ordered ion distributions will occur for ${\cal F}=1/2$,
as opposed to the neutral DX picture employed here, where ordering increases
monotonically as ${\cal F}$ decreases \cite{Heiblum_private}.

\begin{figure}[hbt]
\setlength{\parindent}{0.0in}
 \begin{minipage}{\linewidth}
\epsfxsize=8cm
\epsfbox{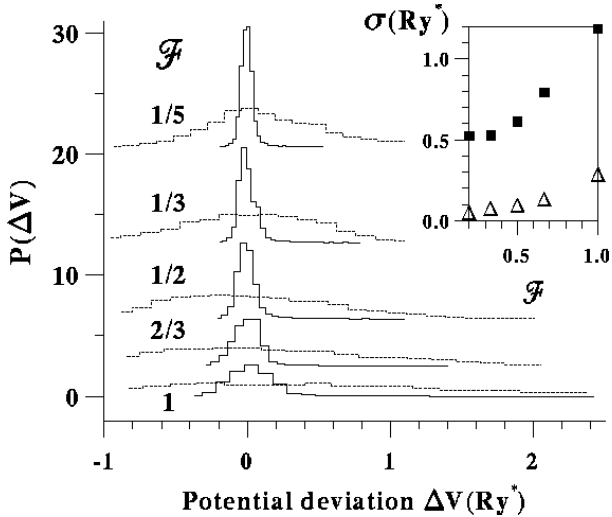}
\vspace*{3mm}
\caption{Histograms of deviation of effective 2D potential from 
quiet dot values at the same $x,y$ point and same gate voltages,
for several ion to donor ratios ${\cal F}$, TF. Solid lines are
statistics for points below Fermi surface, dashed lines, showing
substantially more variation, above. ${\cal F}=1$ is completely
random (disordered) case. Distributions uniformly asymmetric,
positive potential deviations from quiet dot case being more
likely, but means are very close to zero. Inset shows standard
deviation of histograms versus ${\cal F}$, triangle below, 
squares above Fermi level. \label{fig5} }
 \end{minipage}
\end{figure}

For these assumptions figure \ref{fig5} indicates that ionic ordering
substantially reduces the potential fluctuations relative to the
completely disordered case, even for relatively large ${\cal F}$. 
Here, using ensembles
of dots with varying ${\cal F}$ we compare the effective 2D potential
with a quiet dot (jellium donor layer) 
at the same gate voltages and same dot electron
number. The distribution of the potential deviation is computed as:
\begin{equation}
P(\Delta V) = \frac{1}{SN^2} \sum_s \sum_{i,j} \delta(\Delta V - 
[V_{\cal F} (x_i,y_j) - V_{qd} (x_i,y_j)])
\end{equation}
where $s$ labels samples (different ion distributions),
typically up to $S=10$, $N$ is the total number of $x$ or $y$ grid points
in the dot ($\sim 50$), and ``qd'' stands for quiet dot. 
The distributions for all ${\cal F}$ are asymmetric (Fig. \ref{fig5}).
Although the means are indistinguishably close to zero, the probability for
large potential hills resulting from disorder is greater than for deep
depressions. Also, the distributions for points above the Fermi surface 
(dashed lines)
are broader by an order of magnitude (in standard deviation) than below, due to 
screening. Finally, saturation as
${\cal F} \rightarrow 0$ (inset Fig. \ref{fig5}) shows that even if the
ions are arranged in a Wigner
crystal (the limiting case at ${\cal F} = 0$), potential
fluctuations would be expected in comparison with ionic jellium.

\begin{figure}[hbt]
\setlength{\parindent}{0.0in}
 \begin{minipage}{\linewidth}
\epsfxsize=8cm
\epsfbox{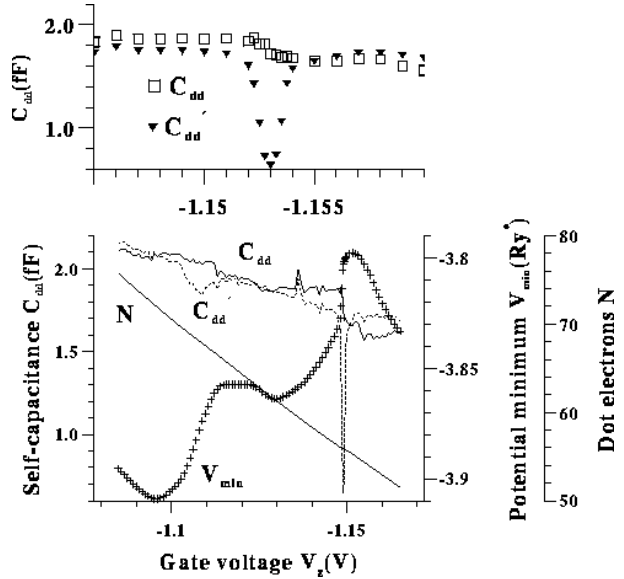}
\vspace*{3mm}
\caption{Dot self-capacitances, equilibrium electron number and 
potential minimum as a function of plunger gate voltage (lower). 
{\it Numerical} uncertainty is indiscernible, so variations
of $C_{dd}$ are real and related to spectrum. $C_{dd}^{\prime}$
calculated using $\Delta E_F$ rather than $\Delta V_{min}$, so
strong anomaly near $-1.15 \; V$ due to rigidity of $N$.
Upper panel: expanded view of capacitances near anomaly;
cf. spectrum, Fig. 9. \label{fig6} }
 \end{minipage}
\end{figure}

The success of the capacitance model in describing experimental results
of charging phenomena in mesoscopic systems has been remarkable \cite{various}. 
For our calculations as well, even the simplest 
formulations for the capacitance tend to produce smoothly varying results
when gate voltages or dot charge are varied. Figure \ref{fig6} shows the trend
of the dot self-capacitances with $V_g$. Also shown are the equilibrium
dot electron number $N$ and the minimum of the dot potential $V_{min}$ as
functions of $V_g$. Note here that $V_{min}$ is the minimum of the 3D
electrostatic potential rather than the effective 2D potential which is
presented elsewhere (such as in Figs. \ref{fig3} and \ref{fig4}). 

That $C_{dd}$ generally decreases as the dot becomes 
smaller is not surprising and has been discussed elsewhere \cite{ep2ds10}.
All three forms of $C_{dd}$ are roughly in agreement giving a
value $\sim 2 \; fF$ (the capacitance as calculated from the
free energy is not shown). The fluctuations result from variations
in the quantized level energies as the dot size and shape are changed by 
$V_g$. Note that {\it numerical} error is indiscernible on the
scale of the figure. The pronounced collapse of $C_{dd}^{\prime}$ 
near $V_g = -1.15 \; V$, which is expanded in the
upper panel, shows the presence of a region where the change
of $N$ with $E_F$ is greatly suppressed. Since the change of $V_{min}$ 
with $E_F$ is similarly suppressed there is no corresponding 
anomaly in $C_{dd}$. Interestingly, the capacitance computed from the free energy
also reveals no deep anomaly. 

The anomaly at $V_g = -1.15 \; V$ and also the fluctuation in 
the electrostatic properties near $-1.1 \; V$ are related to a
shell structure in the spectrum which we discuss below. 

A frequently encountered model for the classical charge
distribution in a quantum dot is the circular conducting disk
with a parabolic confining potential \cite{Shikin,Chklovskii}.
It can be shown (solving, for example, Poisson's equation
in oblate spheroidal coordinates) 
that for such a model the 2D charge distribution
in the dot goes as
\begin{equation}
n(r) = n(0)(1-r^2/R^2)^{1/2} \label{eq:circ}
\end{equation}
where $R$ is the dot radius and $n(0)=3N/2 \pi R^2$ is the 
density at the dot center. 
The ``external'' confining potential is assumed to
go as $V(r)=V_0 + kr^2/2$ and $R$ is related to $N$
through 
\begin{equation}
R = \frac{3 \pi}{4} \frac{e^2}{\kappa k} N
\end{equation}
where $\kappa$ is the dielectric constant \cite{Shikin}. 

To justify this model, the authors of Ref. \cite{Shikin}
claim that the calculations of Kumar {\it et al.} \cite{Kumar}
show that ``the confinement...has a nearly parabolic
form for the external confining potential ({\it sic}).''
This is incorrect. What Kumar {\it et al.}'s calculations shows
is that (for $N \stackrel{<}{\sim} 12$) the {\it self-consistent} potential,
which includes the potential from the electrons themselves,
is approximately cutoff parabolic. The {\it external} confining 
potential, as it is used in Ref. \cite{Shikin},
would be that produced by the donor layer charge and the 
charge on the surface gates only. We introduce a simple model
(see III.B.1 below) wherein this confining potential charge is replaced
by a circular disk of positive charge whose
density is fixed by the doping density and whose radius is determined by
the number of electrons {\it in the dot}. The gates can be 
thought of as merely cancelling the donor charge outside that
radius. The essential point, then, is this: adding 
electrons to the dot decreases the (negative) charge on the gates and
therefore increases the radius. One can make the assumption, as in Ref.
\cite{Shikin}, that the external potential is parabolic, but it is a 
mistake to treat that parabolicity, $k$, as independent of $N$.

\begin{figure}[hbt]
\setlength{\parindent}{0.0in}
 \begin{minipage}{\linewidth}
\epsfxsize=8cm
\epsfbox{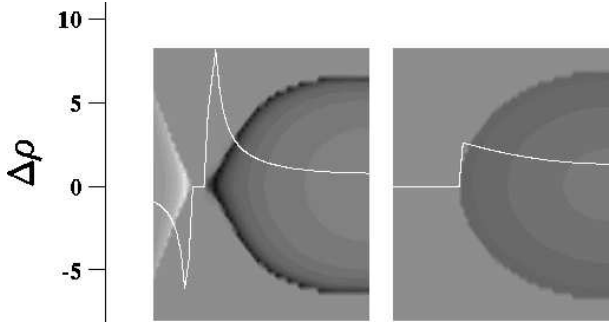}
\vspace*{3mm}
\caption{Grey scale of density change as Fermi energy in dot is raised
relative to leads, Thomas-Fermi (TF). Total change in $N$ about $1.4$
electrons. Screening charge, white region, in leads is positive.
White curve gives profile along line bisecting dot, scaled
to average change of $N$ per unit area. Right panel
shows model of Ref. $~^{44}$ where confining potential
has fixed parabolicity. Note that this model drastically
underestimates degree to which charge is added 
to perimeter. \label{fig7} }
 \end{minipage}
\end{figure}

This is illustrated in 
figure \ref{fig7} where we have plotted contours for the {\it change} in 
the 2D density, as $E_F$ is incrementally increased, 
as determined self-consistently (Thomas-Fermi everywhere, left panel)
and as determined from Eq. \ref{eq:circ}. The white curves display
the density change profiles across the central axis of the dot.
The total change in $N$ is the same in both cases, but
clearly the model of Eq. \ref{eq:circ} underestimates
the degree to which new charge is added mostly to the
perimeter. 

Recently the question of charging energy renormalization via
tunneling as the conductance $G_0$ through a QPC 
approaches unity has received much attention \cite{Matveev2,Halperin,Kane}.
In a recent experiment employing two dots in series a 
splitting of the Coulomb oscillation peaks has been observed
as the central QPC (between the two dots) is lowered
\cite{Westervelt}. Perturbation theory for small $G_0$
and a model which treats the decaying channel between
the dots as a Luttinger liquid for $G_0 \rightarrow 1 \, (e^2/h)$
lead to expressions for the peak splitting which is 
linear in $G_0$ in the former case and goes
as $(1-G_0)ln(1-G_0)$ in the latter case.

\begin{figure}[hbt]
\setlength{\parindent}{0.0in}
 \begin{minipage}{\linewidth}
\epsfxsize=8cm
\epsfbox{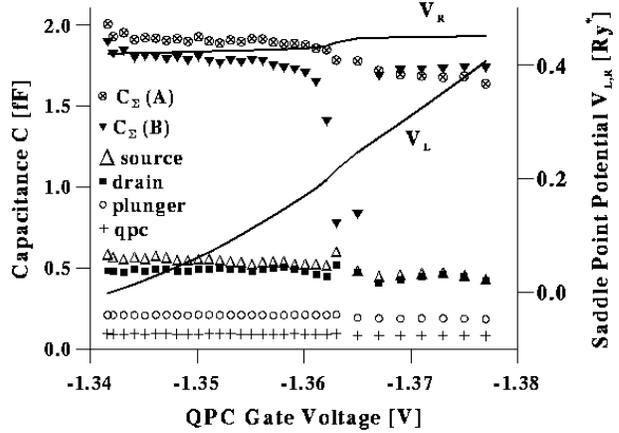}
\vspace*{3mm}
\caption{Variation of dot capacitances with QPC voltage (gates
$1$ and $4$ in figure 1). Solid
lines for $V_{L(R)}$ are effective 2D potential for
left (right) saddle point (right hand scale). 
$C_{\Sigma}(A)$ and $C_{\Sigma}(B)$
are dot self-capacitances (cf. Fig. 5) computed using $\Delta V_{min}$
and $\Delta E_F$ respectively. ``Source'' is (arbitrarily)
outside {\it left}
saddle point. Note that $V_L$ goes practically to zero  
but the dot capacitance to the source only marginally
increases relative to dot to drain capacitance. Capacitances for QPC
and plunger are for a single finger only in each case.
Anomaly related to dot reconstruction also visible here as
QPC voltage is changed. \label{fig8} }
 \end{minipage}
\end{figure}

A crucial assumption of the model, however, is that the
``bare'' capacitance, specifically that between the
dots $C_{d1-d2}$, remains approximately independent 
of the height of the QPC, even when an open channel 
connects the two dots. Thus the mechanism of the peak
splitting is assumed to be qualitatively different
from a model which predicts peak splitting entirely
on an electrostatic basis when the inter-dot capacitance
increases greatly \cite{Ruzin}. The independence
of $C_{d1-d2}$ from the QPC potential is plausible insofar
as most electrons, even when a channel is open, are
below the QPC saddle points and hence localized on either
one dot or the other. Further, if the screening length
is short and if the channel itself does not accommodate
a significant fraction of the electrons, there is
little ambiguity in retaining $C_{d1-d2}$ to
describe the gross electrostatic interaction of the dots,
even when the dots are {\it connected} at the Fermi level.

In figure \ref{fig8} we present evidence for this theory by
showing the capacitance between a dot and the {\it leads} 
as the QPC voltage is reduced. In the figure $V_{L(R)}$
is the effective 2D potential of the left (right)
saddle point as the left QPC gate voltages $V_{QPC}$ only are varied. The
dot is nearly open when the QPC voltages (both pins on the left) reach
$\sim -1.34 \; V$. The results here use the full quantum
mechanical solution (without the LDA exchange-correlation energy),
however the electrons in the lead continue to be treated
with a 2D TF approximation. The dot ``reconstruction''
seen in figure~\ref{fig5} is visible 
\begin{figure}[hbt]
\setlength{\parindent}{0.0in}
 \begin{minipage}{\linewidth}
\epsfxsize=8cm
\epsfbox{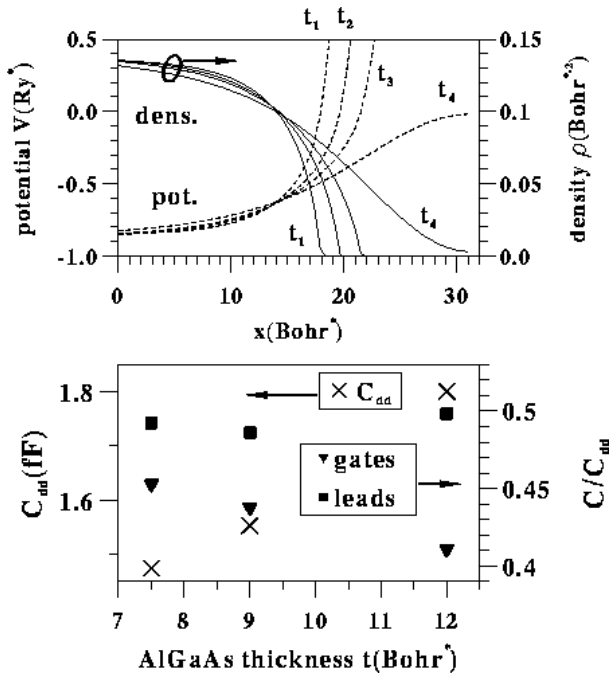}
\vspace*{3mm}
\caption{$AlGaAs$ thickness dependence of capacitances (lower).
Self-capacitance decreases as gates get closer to 2DEG. Upper panel
shows that, for smaller $t$ the potential confinement is
steeper and charge more compact, hence smaller $C_{dd}$. 
$t_1=5.25, \, t_2=7.5, \, t_3=9,$ and $t_4=12 \, a_B^*$.
Relative capacitance from dot to gates and leads fairly
insensitive to $t$. \label{fig9} }
 \end{minipage}
\end{figure}
here also around $V_{QPC}=-1.365 \; V$.
Note that the right saddle point is sympathetically affected when
we change this left QPC. While the effect is faint, $\sim 5 \%$ of the
change of the left saddle, 
the sensitivity of tunneling to saddle point voltage (see also below)
has  resulted in this kind of cross-talk being problematical
for experimentalists. The figure also shows that the 
capacitance between the dot and one lead exceeds that to
a (single) QPC gate or even to a plunger gate. However the
most important result of the figure is to show that
the dot to lead capacitance is largely insensitive to 
QPC voltage. When the left QPC is as closed as 
the right ($V_{QPC} \sim -1.375 \; V$) the
capacitances to the source and drain are equal.
But even near the open condition the capacitance to
the left lead (arbitrarily the ``source'') only
exceeds that to the drain (which is still closed)
minutely. Therefore the assumptions of a ``bare'' capacitance
which remains constant even as contact is made with a lead
(or, in the experiment, another dot) seems to be very well
founded.

As noted above, the interaction between a gate and the
2DEG depends upon the distance of the gates from the 2DEG,
i.e., the $AlGaAs$ thickness $t$. In figure \ref{fig9} we show that,
as we decrease $t$, simultaneously changing the
gate voltages such that $N$ and the saddle point potentials
remain constant, the total dot
capacitance also decreases, but the distribution
of the dot capacitance between leads, gates and (not shown)
back gate change only moderately. That gates closer to the 2DEG 
plane should produce dots of lower capacitance is made
clear in the upper panel of the figure, which shows the
potential and density profile (using TF) near a depletion
region at the side of the dot
at varying $t$ and constant gate voltage. For smaller $t$
the depletion region is widened but the density 
achieves its ungated 2DEG value (here $0.14 \; a_B^{* \, -2}$)
more quickly; a potential closer to hard walled is realized.
In the presence of stronger confinement the capacitance
decreases 
and the charging energy increases.

The profile of the tunnel barriers and the barrier penetration
factors are also dependent on $t$. However we postpone
a discussion of this until the section on tunneling
coefficients.

\subsection{Spectrum}

The bulk electrostatic properties of a dot are, to first
approximation, independent of whether a Thomas-Fermi
approximation is used or Schr\"{o}dinger's equation
is solved. A notable exception to this is the
fluctuation in the capacitances. Figure \ref{fig10} shows
the plunger gate voltage dependence of the energy
levels. The Fermi level of the dot is kept constant
and equal to that of the leads (it is the energy zero). 
Hence as the gate voltage
increases (becomes less negative) $N$ increases.

Since the QPCs lie along the $x$-axis, the dot is never
fully symmetric with respect to interchange of $x$ and 
$y$, however the most symmetric configuration occurs for
$V_g \sim -1.16 \; V$, towards the right side of the plot.
The levels clearly group into quasi-shells with gaps between.
The number of states per shell follows the degeneracy of a 2D parabolic
potential, i.e. 1,2,3,4,... degenerate levels per shell
(ignoring spin). There is a pronounced
tendency for the levels to cluster at the Fermi surface,
here given by $E=0$, which we discuss below. 

\subsubsection{Shell structure}

Shell structure in atoms arises from the approximate 
constancy of individual electron
angular momenta, and degeneracy with respect to $z$-projection.
Since in two dimensions the angular momentum $m$ is fixed in 
the $z$ (transverse) direction, the isotropy of space is broken
and the only remaining manifest degeneracy, and this only for
azimuthally symmetric dots, is with respect to $\pm z$.
A two dimensional parabolic potential, in the absence of
magnetic field, possesses an accidental degeneracy for
which a shell structure is recovered. 

We have shown above that modelling a quantum dot as a 
classical, conducting layer in an {\it external} parabolic potential
$kr^2/2$, where $k$ is independent of the number of electrons
in the dot, ignores the image charge in the surface gates
forming the dot and therefore fails to properly describe
the evolving charge distribution as electrons are added to the dot.
A more realistic model, which {\it explains} the approximate parabolicity
of the {\it self-consistent} potential, and hence the
apparent shell structure, is illustrated in figure \ref{fig11}.
\begin{figure}[hbt]
\setlength{\parindent}{0.0in}
 \begin{minipage}{\linewidth}
\epsfxsize=8cm
\epsfbox{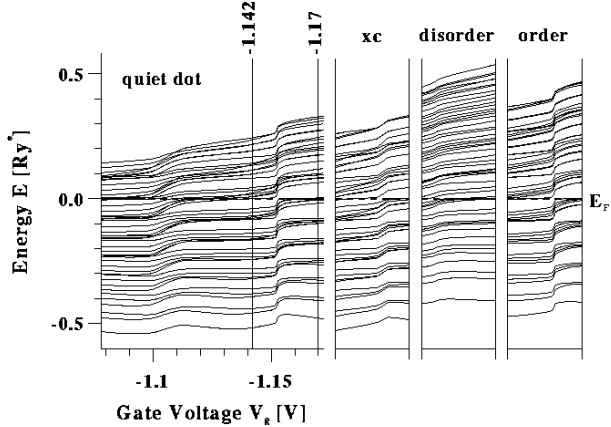}
\vspace*{3mm}
\caption{Electronic spectrum showing level grouping into
shells for quiet dot (Hartree), quiet dot with LDA
exchange-correlation, disordered sample ${\cal F}=1$ and
ordered sample ${\cal F}=1/5$. Range of gate voltage in latter
three is from $V_g = -1.142$ to $-1.17 \; V$. \label{fig10} }
 \end{minipage}
\end{figure}

The basic electrostatic structure of a quantum dot,
in the simplest approximation, can be represented by two circular disks,
of radius $R$ and
homogeneous charge density $\sigma_0$, separated by a distance $a$.
The positive charge outside $R$ is assumed to be cancelled by the 
surface gates. This approximation will be best for
surface gates very close to the donor layer (i.e. small $t$). 
Larger $AlGaAs$ thicknesses will require a non-abrupt termination
\begin{figure}[hbt]
\setlength{\parindent}{0.0in}
 \begin{minipage}{\linewidth}
\epsfxsize=8cm
\epsfbox{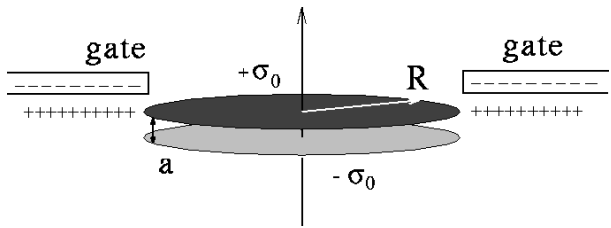}
\vspace*{3mm}
\caption{Schematic for simple two charge disk model of quantum dot.
Positive charge outside radius $R$ taken to be uniformly 
cancelled by gates, electric charge in 2DEG mirrors positive
charge. Resultant radial potential in 2DEG plane, Eq. 15,
dominated by parabolic term inside $R$. \label{fig11} }
 \end{minipage}
\end{figure}
of the positive charge. In either case, the electronic charge is assumed
in the classical limit to screen the background charge as nearly
as possible. This is similar to the postulate in which wide parabolic
quantum wells are expected to produce approximately homogeneous layers
of electronic charge \cite{parabola}.

A simple calculation for the radial potential (for $a<R$)
in the electron layer ($z=0$)
gives, for the first few terms:
\begin{eqnarray}
& \phi(r)= \frac{2 Ne}{\kappa R} [\sqrt{1-a/R} - 1 +
\frac{3}{8} \frac{a^2}{R^2} \frac{r^2}{R^2}  \nonumber \\
& - \frac{15}{32} \frac{a^4}{R^4} \frac{r^2}{R^2} + 
\frac{45}{128} \frac{a^2}{R^2} \frac{r^4}{R^4} + \cdots ] \label{eq:phi}
\end{eqnarray}
where $Ne = \pi R^2 \sigma_0$ and $\kappa$ is the background dielectric
constant. While the coefficient of the quartic term is
comparable to that of the parabolic term, the dependences are scaled
by the dot radius $R$. Hence, the accidental degeneracy of
the parabolic potential is broken only by coupling via the 
quartic term near the dot perimeter. This picture clearly
agrees with the full self-consistent results wherein the 
parabolic degeneracy is observed for low lying states and 
a spreading of the previously degenerate states occurs nearer
to the Fermi surface.

Comparison (not shown) of the potential computed from Eq. \ref{eq:phi} and 
the radial potential profile (lowest curve, Fig. \ref{fig3}b) from the
full self-consistent structure, shows 
good agreement for overall shape. However the former
is about $25 \%$ smaller (same $N$) indicating that the sharp cutoff
of the positive charge is, for these parameters, too extreme. 
However Eq. \ref{eq:phi}
improves for larger $N$ and/or smaller $t$.

\begin{figure}[hbt]
\setlength{\parindent}{0.0in}
 \begin{minipage}{\linewidth}
\epsfxsize=8cm
\epsfbox{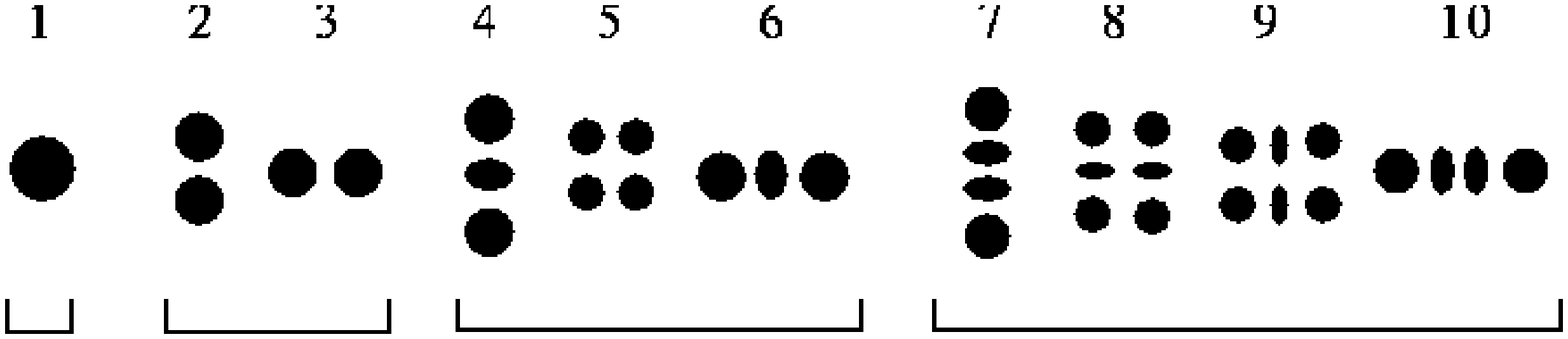}
\vspace*{3mm}
\caption{Schematic showing the first ten levels of quiet dot.
Shell structure consistent with $n+m=$ constant, where
$n$ and $m$ are nodes in $x$ and $y$. Lower energy states
show rectangular symmetry. \label{fig12}}
\vspace*{6mm}
\epsfxsize=8cm
\epsfbox{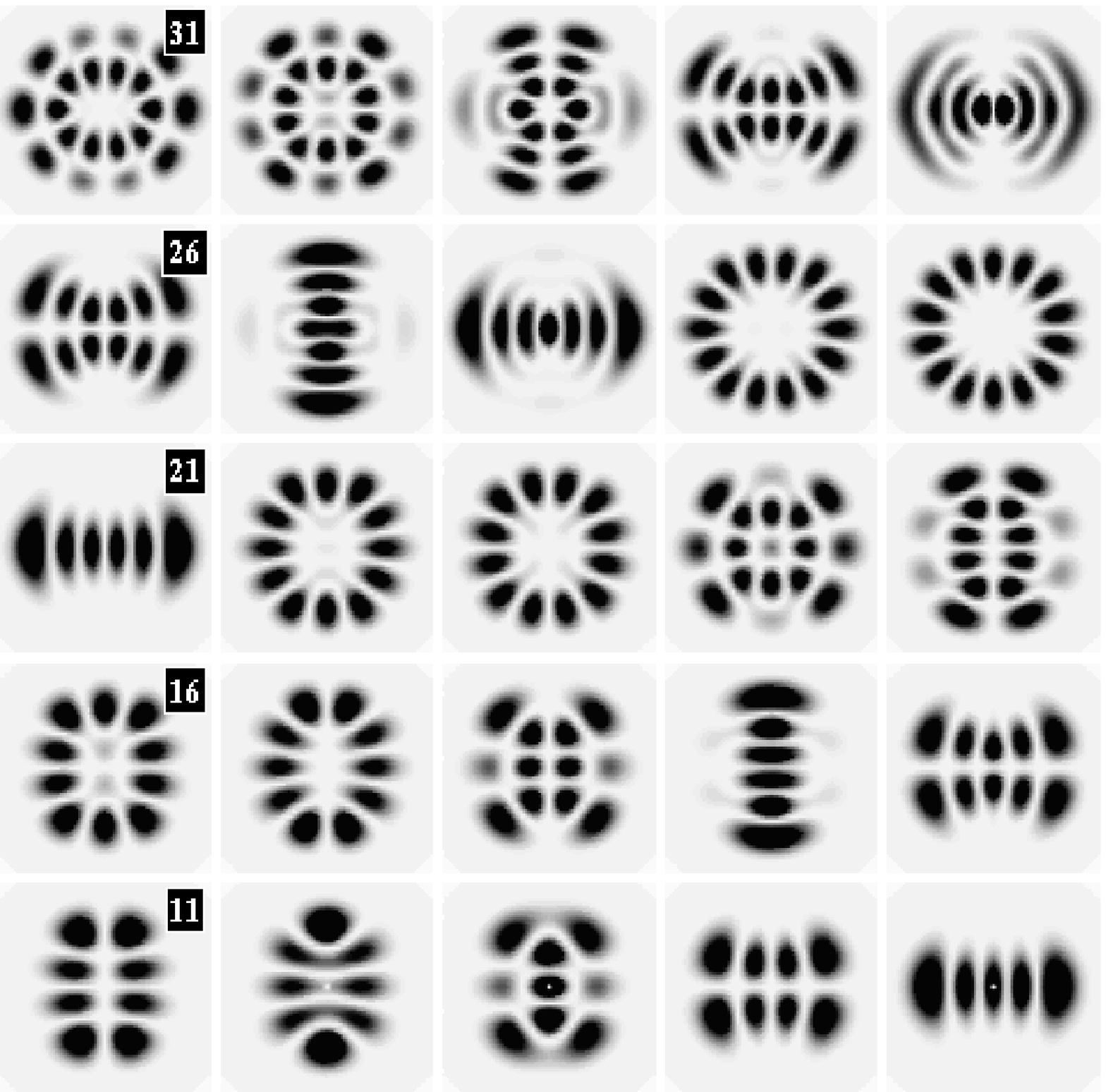}
\vspace*{3mm}
\caption{Levels $11$ through $35$ (each spin degenerate) of
quiet dot, Hartree. Circular symmetry increases with energy. States
elongated in $x$ (horizontal) most connected to leads. \label{fig13}}
 \end{minipage}
\end{figure}

The wavefunction moduli squared associated with the Fig. \ref{fig10}
quiet dot levels for $V_g \sim -1.16 \; V$, $N \approx 54$ are shown
schematically for levels $1$ through $10$ in figure \ref{fig12}, and
for levels $11$ through $35$ in figure \ref{fig13}. 

The lowest level in a shell is, for the higher shells,
typically the most circularly symmetric. When the last
member of a shell depopulates with $V_g$ the inner
shells expand outward, as can be seen near $V_g = -1.15 \; V$
(Fig. \ref{fig10})
where level $p=29$ depopulates. Since to begin filling a new shell
requires the inward compression of the other shells
and hence more energy, the 
capacitance decreases in a step when a shell is depopulated.
The shell structure should have two distinct
signatures in the standard (electrostatic) Coulomb
oscillation experiment \cite{various}. 
First, since the self-capacitance drops appreciably
(figure \ref{fig6}) when the last member of a shell depopulates,
here $N$ goes from $57$ to $56$, a concomitant discrete rise
in the activation energy in the minimum between Coulomb
oscillations can be predicted.
Second, envelope modulation of peak heights \cite{RComm}
occurs when excited dot states are thermally accessible
as channels for transport, as opposed to the $T=0$ case where
the only channel is through the first open state
above the Fermi surface (i.e. the $N+1^{st}$ state).
When $N$ is in the middle of a shell of closely spaced, spin
degenerate levels, the entropy of the dot, $k_B ln \Omega $,
where $\Omega$ is the number of states accessible to the
dot, is sharply peaked. For example, for six electrons
occupying six spin degenerate levels (i.e. twelve altogether)
all within $k_B T$ of the Fermi surface, the number of channels 
available for transport is $924$. For eleven electrons in the
shell, however, the number of channels reduces to $12$.
Consequently, minima and peaks of envelope modulation
(see also figure \ref{fig22} below)
of CB oscillations which are frequently observed are clear
evidence of level bunching, if not an organized shell structure.

Recently experimental evidence has accumulated for the
existence of a shell structure as observed by
inelastic light scattering \cite{Lockwood} and via Coulomb oscillation
peak positions in transport through
extremely small ($N \sim 0-30$) vertical
quantum dots \cite{Tarucha}. Interestingly, a {\it classical} 
treatment, via Monte-Carlo molecular dynamics simulation \cite{Peeters}
also predicts a shell structure. Here, 
the effect of the neutralizing positive background
are assumed to produce a parabolic confining potential.
A similar assumption is made in Ref. \cite{Akera} which
analyzes a vertical structure similar to that of Ref. \cite{Tarucha}.
We believe that continued advances in fabrication will result
in further emphasis on such invariant, as opposed to merely statistical,
properties of dot spectra.

As noted above, there is a strong tendency for levels at the
Fermi surface to ``lock.'' Such an effect has been described by
Sun {\it et al.} \cite{Sun} in the case of subband levels for
parallel quantum wires.
In dots, the effect
can be viewed as electrostatic pressure on the 
individual wavefunctions thereby shifting
level energies in such a way as to produce level
{\it occupancies} which minimize the
total energy. Insofar as a given set of level occupancies
is electrostatically most favorable, level locking 
is a temperature dependent effect which increases as
$T$ is lowered. This self-consistent modification of
the level energies can also be viewed as an excitonic
correction to excitation energies.

The difference between the cases of a quantum dot and that
of parallel wires is one of localized versus extended systems.
It is well known that, unlike Hartree-Fock theory, wherein
self-interaction is completely cancelled since the direct and
exchange terms have the same kernel $1/|{\bf r} - {\bf r^{\prime}}|$,
in Hartree theory and even density functional theory in
the LDA, uncorrected self-interaction remains \cite{Perdew}.
While it is reasonable to expect that excited states will have their energies
corrected downward by the remnants of an excitonic effect, we expect
that LDA and especially Hartree calculations will generally overestimate this
tendency to the extent that corrections for self-interaction are
not complete. 

The panel labelled ``xc'' in figure \ref{fig10} illustrates the preceding point.
In contrast to the large panel (on the left) these results have
had the XC potential in LDA included. The differences
between Hartree and LDA are generally subtle, but here the clustering
of the levels at the Fermi surface is clearly mitigated by the
inclusion of XC. The approximate parabolic degeneracy is
evidently not broken by LDA, however, and the shell structure
remains intact.
Similarly for xc, the capacitances
also show anomalies near the same gate
voltages, where shells depopulate, as in figure \ref{fig5}, 
which is pure Hartree.

The two remaining panels in figure \ref{fig10} illustrate the effects of
disorder and ordering in the donor layer (XC not included). 
As with the ``xc'' panel, $V_g$ is varied between $-1.142$ and $-1.17 \; V$.
The ``disorder'' panel represents a single fixed distribution of
ions placed at random in the donor layer as discussed above.
Similarly, the ``order'' panel represents a single ordered distribution
generated from a random distribution via the Monte-Carlo simulation
\cite{BR2,ISQM2}; here ${\cal F} = 1/5$ (cf. two panels of Fig. \ref{fig4}).

The shell structure, which is completely destroyed for fully
random donor placement (see also Fig. \ref{fig15}), is almost perfectly 
recovered in the
ordered case. In both cases the energies are uniformly shifted upwards
relative to the quiet dot by virtue of the discreteness of donor
charge (cf. also discussion of Fig. \ref{fig5} above). Closer examination
of the disordered spectrum shows considerably more level repulsion
than the other cases. 

\begin{figure}
\setlength{\parindent}{0.0in}
 \begin{minipage}{\linewidth}
\epsfxsize=8cm
\epsfbox{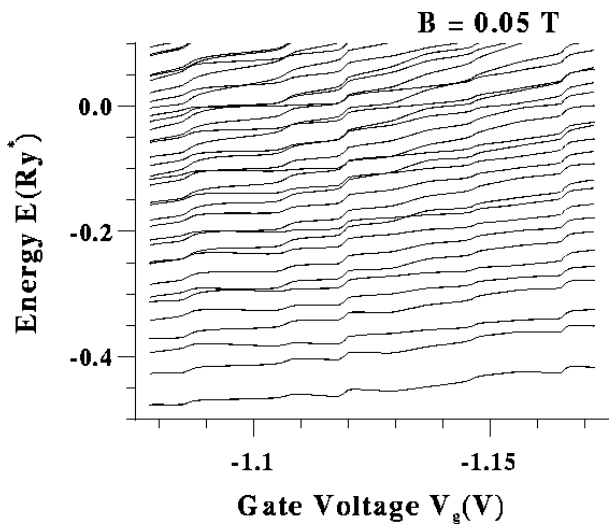}
\vspace*{3mm}
\caption{$V_g$ dependence at fixed $B$ ($0.05 \; T$)
of level energies, quiet dot. Multiple 
re-constructions seen as levels depopulate.
Homogeneous level spacing related
to uniformity of Coulomb oscillation peak heights in a magnetic 
field. \label{fig14} }
 \end{minipage}
\end{figure}

\begin{figure}[hbt]
\setlength{\parindent}{0.0in}
 \begin{minipage}{\linewidth}
\epsfxsize=8cm
\epsfbox{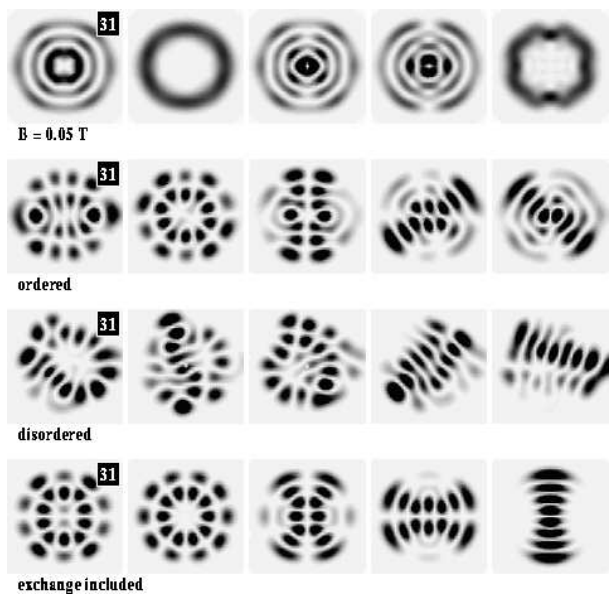}
\vspace*{3mm}
\caption{Levels $31$ through $35$ for (from bottom) quiet dot with LDA
for XC, Hartree for disordered dot, Hartree for ordered dot 
${\cal F}=1/5$ and $B=0.05 \; T$. XC changes ordering
of some levels, but has very little influence on states. 
Ordered case recovers much of quiet dot symmetry. Small $B$
changes states altogether. \label{fig15}}
 \end{minipage}
\end{figure}

The application of a small magnetic field, roughly a single
flux quantum through the dot, has a dramatic impact on both
the spectrum, figure \ref{fig14}, and the wave functions, figure \ref{fig15}, 
top. The magnetic field dependence of the levels (not shown) up to $0.1 \; T$
exhibits shell splitting according to azimuthal quantum number as
well as level anti-crossing. By $0.05 \; T$ level spacing (Fig. \ref{fig14})
is substantially more uniform than $B=0$,
Fig. \ref{fig10}. Furthermore, while 
the $B=0$ quiet dot displays reconstruction due to the depopulation of 
shells at $V_g \approx -1.15$ and $-1.1 \; V$, 
the $B=0.05 \; T$ results show a similar pattern, a step in
the levels, repeated
many times in the same gate voltage range. The physical meaning of this
is clear. The magnetic field principally serves to remove the
azimuthal dependence of the mod squared of the wave functions 
(Fig. \ref{fig15}). In a magnetic field, the 
states at the Fermi surface also tend to be at the dot perimeter.
Depopulation of an electron in a magnetic field, like depopulation of
the last member of a shell for $B=0$, therefore
removes charge from the perimeter of the dot and a self-consistent 
expansion of the remaining states outward occurs. 

\subsection{Statistical properties}

\subsubsection{Level spacings}

The statistical spectral properties of quantum systems
whose classical Hamiltonian is chaotic are believed to
obey the predictions of random matrix theory (RMT) \cite{Andreev}. 
Arguments for this conjecture however invariably treat
the Hamiltonian as a large finite matrix with averaging
taken only near the band center. Additionally, an 
often un-clearly stated assumption is that the system in question can
be treated {\it semi-classically}, that is, in some sense the action is
large on the scale of Planck's constant and the wavelength {\it of all
relevant states} is short on the scale of the system size.
Clearly, for small quantum dots these assumptions are violated.

RMT predictions apply to level spacings $S$
and to transition amplitudes (for the ``exterior problem,''
level widths $\Gamma$) \cite{Brody}. RMT is also
applied to scattering matrices in investigations of
transport properties of
quantum wires \cite{Slevin}. 
Ergodicity for chaotic systems is the claim that variation of
some external parameter $X$ will sweep the Hamiltonian rapidly through 
its entire Hilbert space, whereupon energy averaging
and ensemble (i.e. $X$) averaging produce identical statistics. 
In our study $X$ is either the set of gate voltages, the
magnetic field or the impurity configuration
and we consider the statistics of the lowest
lying $45$ levels (spin is ignored here). Care must also be taken
in removing the secular variations of the spacings or widths
with energy, the so-called unfolding. 

\begin{figure}[hbt]
\setlength{\parindent}{0.0in}
 \begin{minipage}{\linewidth}
\epsfxsize=8cm
\epsfbox{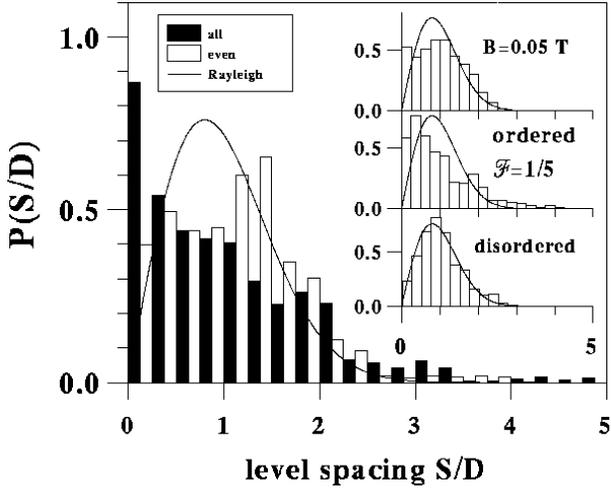}
\vspace*{3mm}
\caption{Histograms of level spacings, normalized to local level spacing.
Dark curve represents Rayleigh distribution. Black bars (main panel)
include all states, white bars only for states that are completely
even under $x$ or $y$ inversion. Insets: disordered panel
recapitulates Rayleigh distribution, both ordered and $B \ne 0$ 
marginally but significantly different. \label{fig16} }
 \end{minipage}
\end{figure}

According to RMT level repulsion leads to statistics of level
spacings which are given by the ``Rayleigh distribution:''
\begin{equation}
P(S)=\frac{\pi S}{2D} exp(-\pi S^2/4 D^2) \label{eq:stat}
\end{equation}
where $D$ is the mean local spacing \cite{Brody,Wigner}.
Figure \ref{fig16} shows the calculated histogram for the level spacings
for the quiet dot as well as for disordered, ordered and
ordered with $B=0.05 \; T$ cases. Statistics are generated from
(symmetrical) plunger gate variation,
in steps of $0.001 \; V$, over a range of $0.1 \; V$, employing the spacings
between the lowest $45$ levels; thus about $4500$ data points.
Deviation from the Rayleigh distribution is evident. An important
feature of our dot is symmetry under inversion through both
axes bisecting 
the dot. It is well known that groups of states which
are
un-coupled will, when plotted together, show a Poisson distribution for
the spacings rather than the level repulsion of Eq. \ref{eq:stat}.
Thus we have also plotted (white bars) the statistics for 
those states which are totally even in
parity. While the probability of degeneracy decreases, 
a $\chi^2$ test shows that the distribution remains 
substantially removed from the Rayleigh form.

In contrast to this, the disordered case shows remarkable
agreement with the RMT prediction. As with the spectrum
in figure \ref{fig10} we use a single ion distribution. However we
also find (not shown) that fixing the gate voltage and varying the random
ion distributions results in nearly the same statistics. 
When the ions are allowed to order the level statistics again
deviate from the RMT model. This is somewhat surprising
since Fig. \ref{fig5} shows that, even for ${\cal F}= 1/5$, the standard 
deviation
of the effective 2D potential below the Fermi surface from the
quiet dot case, $\sim 0.05 \; Ry^*$, is still substantially greater than 
the mean level spacing $\sim 0.02 \; Ry^*$. We have recently
shown that, as ${\cal F}$ goes from unity to zero, a continuous transition
from the level repulsion of Eq. \ref{eq:stat} to a Poisson
distribution of level spacings results \cite{NanoMes}.
Finally, the application of
a magnetic field strong enough to break time-reversal symmetry
clearly reduces the incidence of very small spacings, but the
distribution is still significantly different from RMT.

\subsubsection{Level widths}

\begin{figure}[hbt]
\setlength{\parindent}{0.0in}
 \begin{minipage}{\linewidth}
\epsfxsize=8cm
\epsfbox{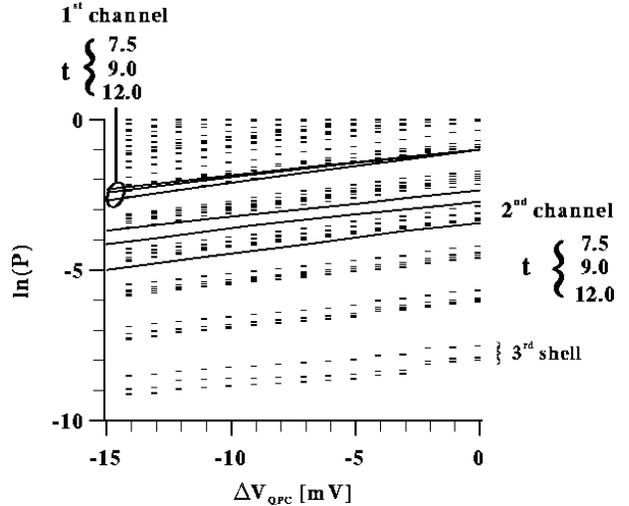}
\vspace*{3mm}
\caption{Barrier penetration factors from classical turning point in lead
to turning point in dot at same energy, as a function
of QPC voltage offset. $P$ evaluated at energies of
states in quiet dot for $AlGaAs$ thickness $t=12 \; a_B^*$. 
Solid lines indicate barrier penetration
at Fermi level. Upper three lines for first channel, $t = 7.5,9.0,12.0 \ a_B^*$
respectively. Lower three lines for second channel, same $t$.
$\Delta V_{QPC}$ zero set such that first channel conducts
equally at the Fermi surface for all $t$. \label{fig17}}
 \end{minipage}
\end{figure}

In Eq. \ref{eq:tun} we defined $W_n(a,b)$ as the barrier penetration factor
from the classically accessible region of the lead to the matching point in
the barrier, for the $n^{th}$ channel. The
penetration factor {\it completely} through the barrier, $P_n \equiv W_n(a,c)$
where $c$ is the classical turning point on the dot side of the barrier,
is plotted as a function of QPC voltage in figure \ref{fig17}. 
$P_n$ is simply the WKB penetration for a given channel with
a given self-consistent barrier profile, and
can be computed at any energy. Here we have computed it
at energies coincident with the dot levels. Therefore the dashes
recapitulate the level structure, spaced now not in energy but in
``bare'' partial width. The {\it actual} width of a level 
depends upon the wave 
function for that state (cf. Eq. \ref{eq:tun}). 
For energies above the barrier $ln(P)=0$.
The solid lines represent $P$ {\it at the Fermi surface} computed for
three different $AlGaAs$ thicknesses $t$ (as in figure \ref{fig9}) and 
for both $n=1$ and $n=2$ (the dashes are computed for $t=12 \; a_B^*$). 
The QPC voltage is given relative to
values at which $P$ for $n=1$ is the same for all three $t$
(hence the top three solid lines converge at $\Delta V_{QPC} = 0$). 

Quite surprisingly $t$ has very little influence on the trend of
$P$ with QPC voltage. Note that the ratio of
barrier penetration between the second and first channels $P_2/P_1$
decreases substantially with increasing $t$ since the saddle
profile becomes wider for more distant gates. Even for $t=7.5 \; a_B^*$
however, penetration via the second channel is about a factor
of five smaller than via $n=1$. 

\begin{figure}[hbt]
\setlength{\parindent}{0.0in}
 \begin{minipage}{\linewidth}
\epsfxsize=8cm
\epsfbox{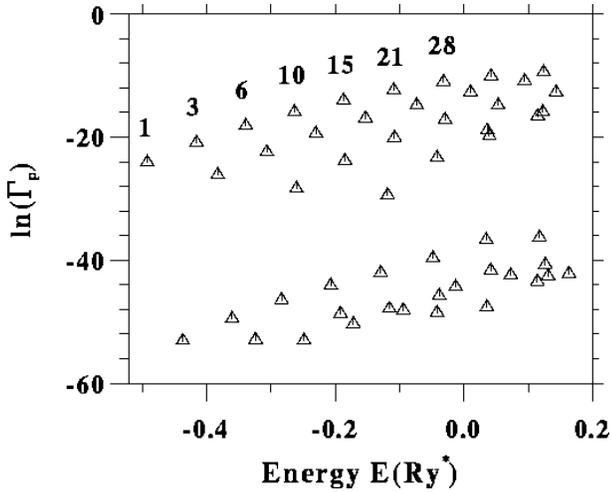}
\vspace*{3mm}
\caption{Partial widths (through first channel) for tunneling to the leads,
quiet dot. Numbers indicate ordinate of wave functions, 
Figs. 11 and 12.
Weakly connected states zero by parity (non-zero only
through numerical error). \label{fig18}}
 \end{minipage}
\end{figure}

Figure \ref{fig18} shows the partial width for tunneling via $n=1$ through the
barrier, now using the full Eq. \ref{eq:tun}, for the quiet dot. 
The barriers here are fairly wide.
While this strikingly coherent structure is quickly destroyed
by discretely localized donors even when donor ordering is allowed,
the pattern is nonetheless highly informative. The principal division
between upper and lower states is based on parity. States which
are odd with respect to the axis bisecting the QPC should in fact
have identically zero partial width (that they don't is evidence of
numerical error, mostly imperfect convergence).
Note that {\it this}
division
is largely preserved for discrete but ordered ions. 
The widest states (largest $\Gamma$) are
labelled with their level index for comparison with their
wave functions in Figs. \ref{fig13} and \ref{fig14}. Comparison shows they represent the states
which are aligned along the direction of current flow. Thus in
each shell there are likely to be a spread of tunneling coefficients,
that is, two members of the same shell will not have the same $\Gamma$.

\begin{figure}[hbt]
\setlength{\parindent}{0.0in}
 \begin{minipage}{\linewidth}
\epsfxsize=8cm
\epsfbox{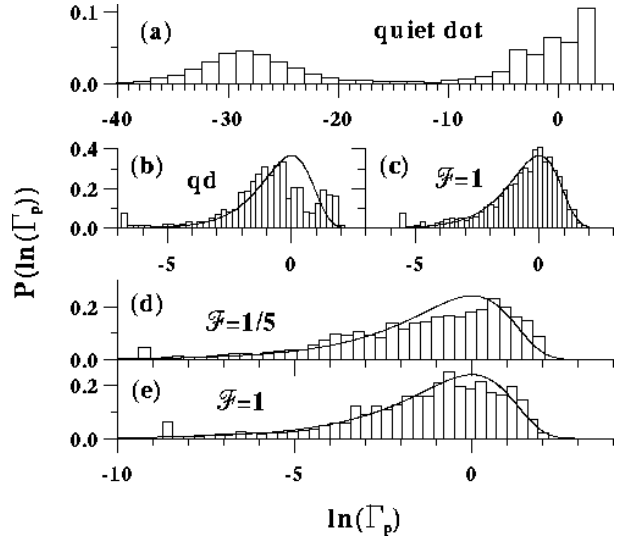}
\vspace*{3mm}
\caption{Statistics of unfolded partial level widths, first channel only,
(a) quiet dot showing large weight near zero due to parity, (b) and (c)
have $B=0.05 \; T$, quiet dot and disordered, respectively. Remnant
of peak at small coupling remains. Dark line represents $\chi_2^2$
distribution predicted by RMT. (d) and (e) are ordered and disordered
with $B=0$. Ordered case differs significantly from Porter-Thomas
distribution plotted in black here. \label{fig19}}
 \end{minipage}
\end{figure}

Statistics of the level partial widths are shown in figure \ref{fig19}, 
here normalized
to their local mean values. While the statistics for the quiet dot
are in substantial disagreement with RMT it is clear that discreteness
of the ion charge, even ordered, largely restores ergodicity.
The RMT prediction, the ``Porter-Thomas'' (PT) distribution, is also plotted.
For non-zero $B$, panels (b) and (c), the predicted distribution 
is $\chi_2^2$ rather than PT. Even the completely disordered case (e)
retains a fraction of vanishing partial width states. Since in our
case the zero width states result from residual reflection symmetry,
it would be interesting to compare the data from references
\cite{Chang} and \cite{Marcus}, which employ nominally symmetric
and non-symmetric dots respectively, to see if the incidence of zero
width states shows a statistically significant difference.

One further statistical feature which we calculate is the
autocorrelation function of the level widths as an external parameter
$X$ is varied:
\begin{equation}
C(\Delta X) =  \\
\frac{\sum_{i,j} \delta \Gamma_i(X_j) 
\delta \Gamma_i(X_j + \Delta X)}
{\sqrt{\sum_{i,j} \delta \Gamma_i(X_j)^2}
\sqrt{\sum_{i,j} \delta \Gamma_i(X_j + \Delta X)^2}} \label{eq:auto}
\end{equation}
where $\delta \Gamma_i(X) \equiv \Gamma_i(X)-\bar{\Gamma}_i(X)$, and
where
$\bar{\Gamma}(X)$ is again the {\it local} average, over levels at fixed
$X$, of the level widths. Note that the sum on $i$ is over levels and the sum on
$j$ is over starting values of $X$. 

\begin{figure}[hbt]
\setlength{\parindent}{0.0in}
 \begin{minipage}{\linewidth}
\epsfxsize=8cm
\epsfbox{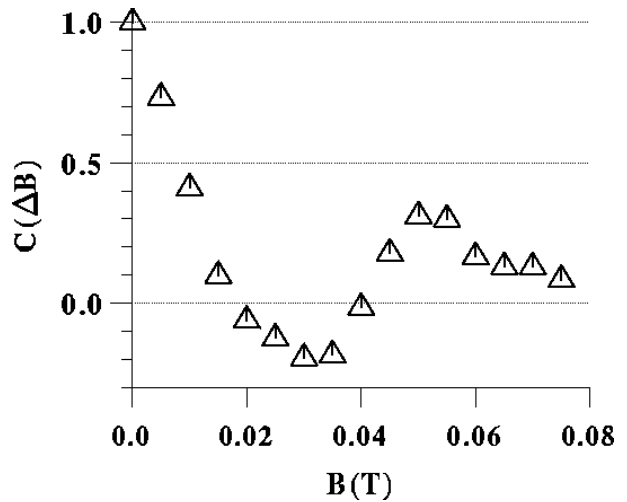}
\vspace*{3mm}
\caption{Autocorrelation function for level partial widths; 
ordered, ${\cal F}=1/5$, averaged over $B$ starting point and
all $45$ levels. Range of $B$ is only $0-0.1 \; T$, so statistics
are weaker to the right. Pronounced anti-correlation near
$0.03 \; T$ in contradiction with RMT. \label{fig20}}
 \end{minipage}
\end{figure}

In figure \ref{fig20} we show the autocorrelation function for varying magnetic
field (cf. Ref. \cite{Marcus}, figure \ref{fig4}). The sample is ordered,
${\cal F}=1/5$.
Our range of $B$ only encompasses
$[0,0.1] \; T$ in steps of $0.005 \; T$, so we have here averaged over 
all levels (i.e. $i=1-45$). The crucial feature, which has been
noted in Refs. \cite{Marcus} and, for conductance correlation
in open dots in \cite{Bird2}, is that the correlation function becomes
negative, in contradiction with a recent prediction based on
RMT \cite{Alhassid}.
Indeed, as noted by Bird {\it et al.} \cite{Bird2}, 
an oscillatory structure seems to emerge in the data.
Comparison with calculation here is hampered since the statistics 
are less good as $B$ increases.

Nonetheless, the RMT prediction is clearly erroneous. 
We speculate that the basis of the discrepancy is in the 
assumption \cite{Alhassid} that $C(\Delta X)=C(-\Delta X)$.
Given this assumption \cite{Ferry2} the correlation 
becomes positive definite. Physically this means that,
regardless of whether $B$ is positive or negative,
the self-correlation of a level width will be independent
of whether $\Delta B$ is positive or negative. This implies
that the level widths should be independent of the 
absolute value of $B$, or any even powers of $B$, at
least to lowest order in $\Delta B/B$. 
For real quantum dot systems this assumption is inapplicable.

\begin{figure}[hbt]
\setlength{\parindent}{0.0in}
 \begin{minipage}{\linewidth}
\epsfxsize=8cm
\epsfbox{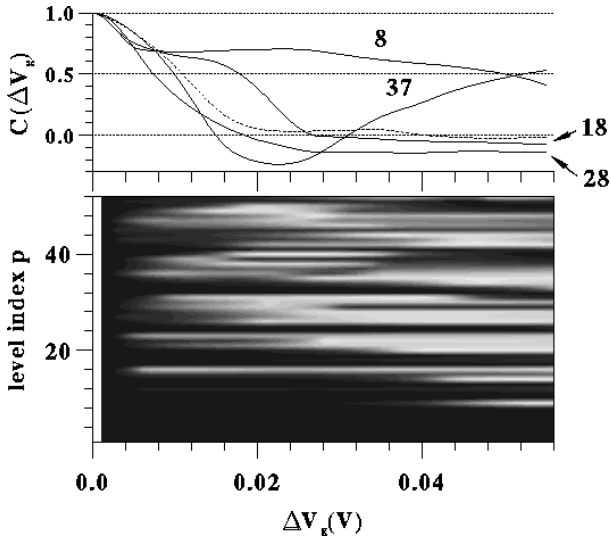}
\vspace*{3mm}
\caption{Autocorrelation function with $V_g$, averaged over groups
of $15$ levels (upper panel). Number indicates center of (contiguous)
range of averaged values. Dashed line is average of all states.
Lower panel is grey scale for autocorrelation of individual 
levels averaged only over $V_g$ starting point. Black is $1.0$
and white is $-1.0$. Data suggests that behaviour of
autocorrelation is sensitive to {\it which} levels are averaged. 
\label{fig21}}
 \end{minipage}
\end{figure}

Similar behaviour is observed with $X$ taken as the (plunger) gate voltage,
for which we have considerably more calculated results, Fig. \ref{fig21}.
The upper panel is the analogue of Fig. \ref{fig20}, only we have broken the
average on levels into separate groups of fifteen levels centered on
the level listed on the figure (e.g., the ``$28$'' denotes a sum
in equ. \ref{eq:auto} of $i=21,35$). the lower panel shows the autocorrelation
as a grey scale for the individual levels (averaging performed
only over starting $V_g$). The very low lying levels, up to $\sim 10$, remain
self-correlated across the entire range of gate voltage. This simply
indicates that the correlation field is level dependent. However,
rather than becoming uniformly grey in a Lorentzian fashion, as
predicted by RMT \cite{Alhassid}, individual
levels tend to be strongly correlated or anti-correlated with their 
original values, and the disappearance of correlation only occurs
as an average over levels.

Again we expect that the explanation for this behaviour lies in the
shell structure. Coulomb interaction prevents states which are nearby in
energy from having common spatial distributions.
Thus in a given range of energy, when one state is strongly connected
to the leads, other states are less likely to be. Further, the ordering
of states appears to survive at least a small amount of disorder in
the ion configuration.

\subsection{Conductance}

The final topic we consider here is the Coulomb oscillation conductance of
the dot. We will here focus on the temperature dependence \cite{RComm},
although statistical properties related to ion ordering are also
interesting.

We have shown in Ref. \cite{RComm} that 
detailed temperature dependence of
Coulomb oscillation amplitudes can be employed as a form
of quantum dot spectroscopy. 
Roughly, in the low $T$ limit the peak heights give the
individual level connection coefficients and, as temperature is
raised activated conductance {\it at the peaks} depends on the
nearest level spacings at the Fermi surface. In this regard
we have explained envelope modulation of peak heights, which
had previously not been understood, as clear
evidence of thermal activation involving tunneling through
excited states of the dot \cite{RComm}.

\begin{figure}[hbt]
\setlength{\parindent}{0.0in}
 \begin{minipage}{\linewidth}
\epsfxsize=8cm
\epsfbox{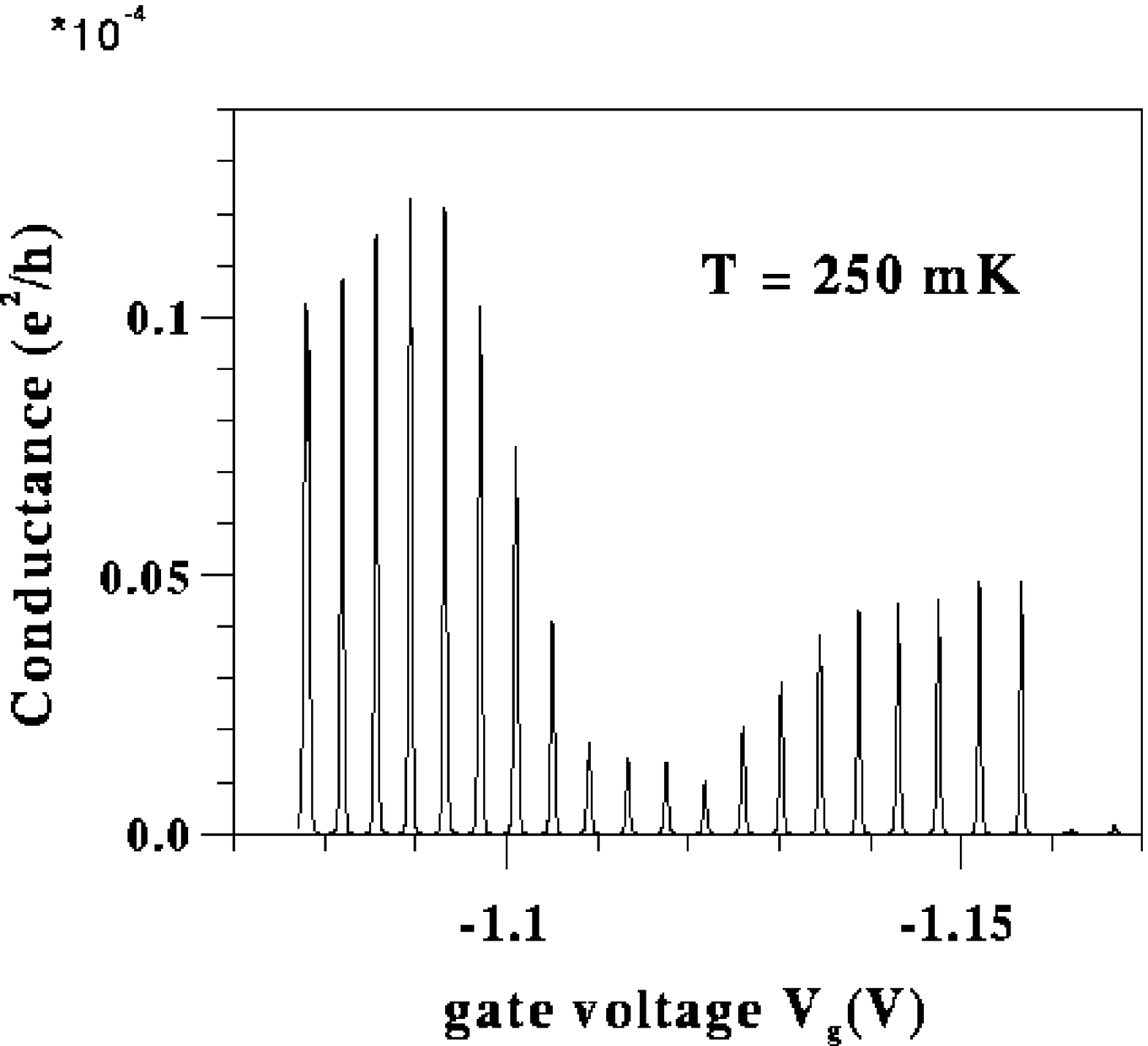}
\vspace*{3mm}
\epsfxsize=8cm
\epsfbox{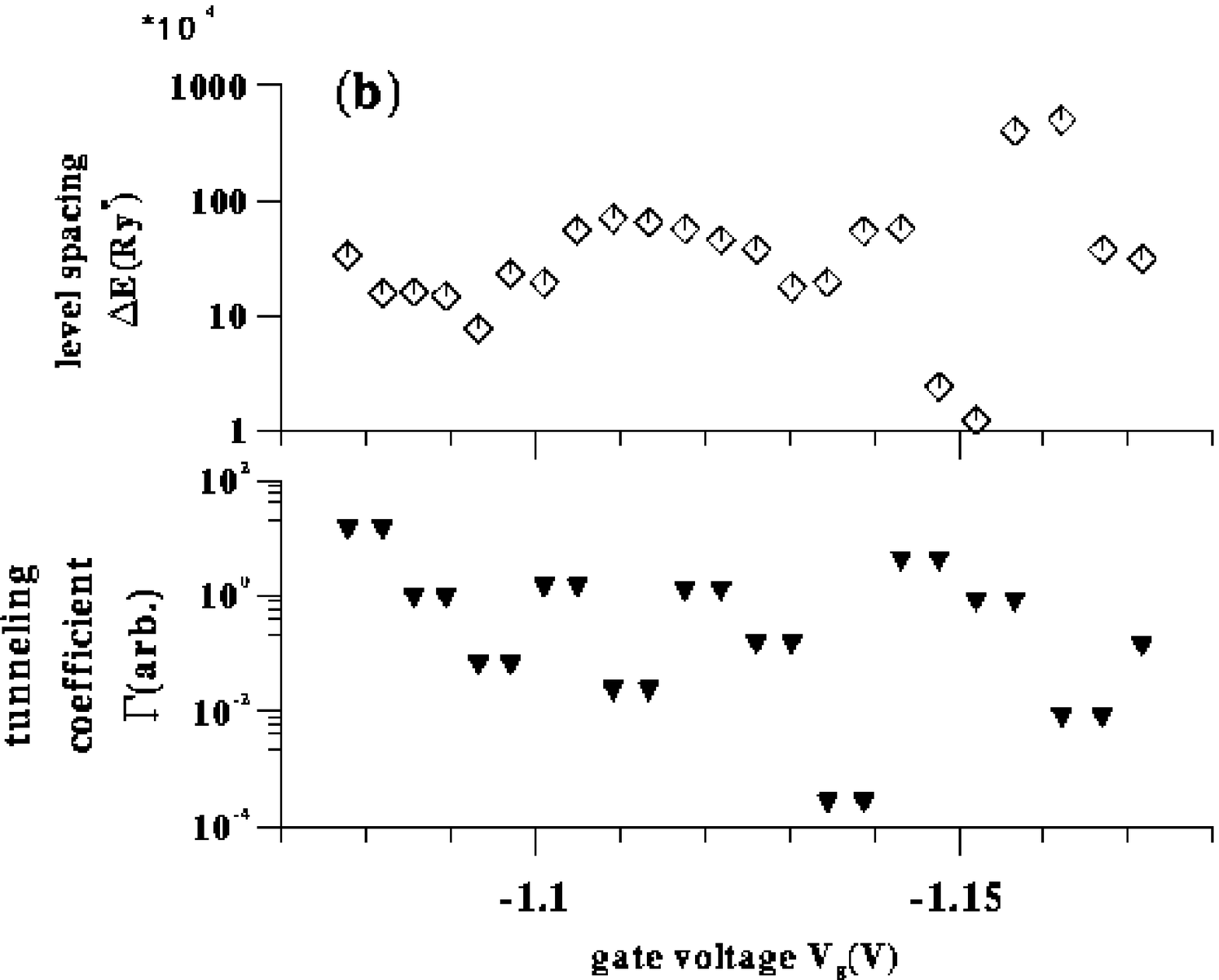}
\vspace*{3mm}
\caption{(a) Conductance versus $V_g$ for ordered dot, $T= 0.25 \; K$.
(b) Fermi surface level spacing and
tunneling coefficient at resonance. Conductance in (a) correlates
somewhat more strongly with smaller level spacing than with larger
$\Gamma$. \label{fig22}}
 \end{minipage}
\end{figure}

Figure \ref{fig22}a 
shows the conductance as a function of plunger gate voltage
for the ordered dot at $T=250 \; mK$. 
Note that the magnitude of the conductance is small because the 
coupling coefficients 
are evaluated with relatively wide barriers for numerical reasons.
Over this range the dot $N$
depopulates from $62$ (far left) to $39$. The level spacings and tunneling
coefficients are all changing with $V_g$. At low temperature a given peak 
height is determined mostly by the coupling to the first empty dot
level ($\Gamma_{N+1}$) and by the spacings between the $N^{th}$ level
and the nearest other level (above or below). The relative importance of
the $\Gamma$'s and the level spacings can obviously vary.
In this example, Figs. \ref{fig22}a and \ref{fig22}b suggest that
peak heights correlate more strongly with the level spacings.
The double envelope coincides with the Fermi level passing through
two shells. In general, the DOS fluctuations embodied in the shell
structure and the
observation (above) that within a shell a spreading of the $\Gamma$'s
(with a most strongly coupled level) results from Coulomb interaction
provide the two fundamental bases of envelope modulation.

Finally, we typically find
that, when peak heights are plotted as a function of
temperature (not shown) some peaks retain
activated conductance down to $T=10 \; mK$. 
Since the dot which we are modelling is small on the scale
of currently fabricated structures, this study suggests that 
claims to have reached the regime where all Coulomb oscillations
represent tunneling through a single dot level are questionable. 

\section{Conclusions}

We have presented extensive data from calculations on the electronic
structure of lateral $GaAs-AlGaAs$ quantum dots, with electron
number in the range of $N=50-100$. Among the principal conclusions which we
reach are the following.

The electrostatic profile of the dot is determined by metal gates
at fixed voltage rather than a fixed space charge. As a consequence of
this the model of the dot as a conducting disk with fixed, 
``external,'' parabolic confinement is incorrect.
Charge added to the dot resides much more at the dot
perimeter than this model predicts. 

The assumption of complete disorder in the donor layer is probably
overly pessimistic. In such a case the 2DEG
electrostatic profile is completely dominated by the ions
and it is difficult to see how workable structures could be
fabricated at all. The presence of even a small degree of
ordering in the donor layer, which can be experimentally
modified by a back gate, dramatically reduces potential
fluctuations at the 2DEG level.

Dot energy levels show a shell structure
which is robust to ordered donor layer ions, though for complete
disorder it appears to break up. The shell
structure is responsible for variations in the capacitance with gate
voltage as well as envelope modulation of Coulomb oscillation peaks.
The claims that Coulomb oscillation data through
currently fabricated lateral quantum dots shows unambiguous
transport through single levels are questionable, though
some oscillations will saturate at a higher temperature than others.

The capacitance between the dot and a lead increases only
very slightly as the QPC barrier is reduced. Thus the electrostatic 
energy between dot and leads is dominated by charge below the Fermi
surface and splitting of oscillation peaks through double dot 
structures \cite{Westervelt} is undoubtedly a result of tunneling.

Finally, chaos is well known to be mitigated in quantum
systems where barrier penetration is non-negligible \cite{Smilansky}.
Insofar as non-inegrability of the underlying classical 
Hamiltonian is being used as the justification for an
assumption of ergodicity \cite{Jalabert} in quantum dots, our
results suggest that further success in comparison with real
(i.e. experimental) systems will occur only when account is
taken in, for example, the level velocity \cite{Alhassid,Simons}, 
of the correlating influences of quantum mechanics.

\acknowledgements

I wish to express my thanks for benefit I have gained in conversations
with many colleagues. These include but are not limited to:
Arvind Kumar, S. Das Sarma, Frank Stern, J. P. Bird, Crispin Barnes,
Yasuhiro Tokura, B. I. Halperin, Catherine Crouch, R. M. Westervelt,
Holger F. Hofmann, Y. Aoyagi, 
K. K. Likharev, C. Marcus and D. K. Ferry. I am also grateful for support
from T. Sugano, Y. Horikoshi, and S. Tarucha. 
Computational support from the Fujitsu VPP500 Supercomputer
and the Riken Computer Center is also
gratefully acknowledged.

\end{multicols}


\begin{references}
\bibitem{square} In other words, since the dot floor is not
flat, even a lithographically square dot, say, with gates
very close to the 2DEG plane would seem to suffer 
rounding of the resulting potential departing appreciably
from the flat bottomed, hard walled square shape envisioned.
\end{references}
\end{document}